\documentclass[aps,prd,preprintnumbers,superscriptaddress,nobibnotes,floatfix,longbibliography,nofootinbib]{revtex4-2}

\pdfoutput=1
\usepackage{amsmath,amsfonts,amssymb,mathrsfs,graphicx,color,longtable}
\usepackage{hyperref}
\usepackage{bm}
\usepackage{array}
\usepackage{color}
\usepackage{enumitem}
\usepackage[explicit]{titlesec}
\usepackage{ulem}
\usepackage[utf8]{inputenc}
\usepackage{float}  
\usepackage{multirow,bigdelim}
\usepackage{blindtext}
\usepackage{rotating}
\usepackage[ddmmyy,24hr]{datetime}

\usepackage[TABBOTCAP,tight]{subfigure}
\newcolumntype{C}[1]{>{\centering\let\newline\\\arraybackslash\hspace{4pt}}m{#1}}

\graphicspath{{plots/}}

\begin{document}

\title{Improved sensitivities of ESS$\nu$SB from a two-detector fit}

\author{F. Capozzi}
\email{francesco.capozzi@univaq.it}
\affiliation{Dipartimento di Scienze Fisiche e Chimiche, Universit\`a degli Studi dell’Aquila, 67100 L’Aquila, Italy}

\author{C. Giunti}
\email{carlo.giunti@to.infn.it}
\affiliation{Istituto Nazionale di Fisica Nucleare (INFN), Sezione di Torino, Via P. Giuria 1, I--10125 Torino, Italy}

\author{C.A. Ternes}
\email{ternes@to.infn.it}
\affiliation{Istituto Nazionale di Fisica Nucleare (INFN), Sezione di Torino, Via P. Giuria 1, I--10125 Torino, Italy}
\affiliation{Dipartimento di Fisica, Universit\`a di Torino, via P. Giuria 1, I–10125 Torino, Italy}

\date{\dayofweekname{\day}{\month}{\year} \ddmmyydate\today, \currenttime}

\begin{abstract}
We discuss the improvement of the sensitivity of ESS$\nu$SB
to the discovery of CP violation and to new neutrino physics
which can be obtained with a two-detector fit of the data
of the near and far detectors.
In particular, we consider
neutrino non-standard interactions
generated by very heavy vector mediators,
nonunitary neutrino mixing, and
neutrino oscillations due to the mixing of the ordinary active neutrinos
with a light sterile neutrino.
\end{abstract}

\maketitle
\tableofcontents


\section{Introduction}
\label{sec:intro}

The European Spallation Source neutrino Super-Beam
(ESS$\nu$SB)~\cite{ESSnuSB:2013dql,Wildner:2015yaa,ESSnuSB:2021azq,Alekou:2022mav}
is a proposed long-baseline neutrino oscillation experiment
with a high intensity neutrino beam produced at the upgraded
ESS facility in Lund (Sweden).
The main goal of ESS$\nu$SB is the search for CP violation in the lepton sector
with a megaton underground Water Cherenkov far detector installed
at a distance of either 360 km or 540 km from the ESS.
A near detector located at a distance of 250 m from the neutrino source will
be used to monitor the neutrino beam intensity and reduce the systematic uncertainties.
Besides CP violation~\cite{Agarwalla:2014tpa,Blennow:2014sja}, the ESS$\nu$SB experiment can
probe new neutrino physics beyond the standard three-neutrino mixing framework
and new neutrino interactions beyond the Standard Model~\cite{Blennow:2014fqa,Blennow:2015nxa,KumarAgarwalla:2019blx,Ghosh:2019zvl,Choubey:2020dhw,Chatterjee:2021xyu,Chatterjee:2022mmu}.

In this paper we discuss the potentialities of a two-detector analysis
of the data of ESS$\nu$SB with the updated
configuration described in Refs.~\cite{ESSnuSB:2021azq,Alekou:2022mav}.
We show the advantages of a two-detector analysis
with respect to an analysis of the data of the far detector alone,
as done in some of the previous studies~\cite{Agarwalla:2014tpa,Blennow:2014sja,KumarAgarwalla:2019blx,Chatterjee:2021xyu}.
We discuss the sensitivity of ESS$\nu$SB to the discovery of CP violation
(Section~\ref{sec:SM}),
to neutrino non-standard interactions (NSI)
generated by very heavy vector mediators
(Section~\ref{sec:NSI}),
to nonunitarity of the neutrino mixing matrix
(Section~\ref{sec:NU}), and
to neutrino oscillations due to the mixing of the ordinary active neutrinos
with a light sterile neutrino
(Section~\ref{sec:3+1}).

\section{Simulation details}
\label{sec:sim}
In this section we describe the simulation details of ESS$\nu$SB. Our simulation of the far detector follows Refs.~\cite{ESSnuSB:2021azq,Alekou:2022mav} using the GLoBES~\cite{Huber:2004ka,Huber:2007ji} files for ESS$\nu$SB, which take into account the ESS$\nu$SB specific fluxes, selection efficiencies, energy smearing and cross sections. The far detector is a water Cherenkov detector with 538~kt fiducial volume at 360~km distance from the source. There is another possible location at 540~km distance from the source. However, it has been shown that the CP sensitivity is slightly worse~\cite{ESSnuSB:2021azq,Alekou:2022mav}, and we will not consider this possibility here. We assume a running time of 5 years in neutrino mode and 5 years in antineutrino mode. In addition to the far detector we simulate a near detector assuming the same energy smearing as for the far detector. The near detector is also a water Cherenkov detector, but placed at 250~m from the source and with a fiducial volume of 1~kt.

In our analysis we include appearance and disappearance channels. Each channel includes background contributions from wrong sign contamination of the beam, flavor misinterpretations and misinterpretations of event topologies. In addition to the considerations of Refs.~\cite{ESSnuSB:2021azq,Alekou:2022mav}, which included only charged current (CC) events, we also simulate a neutral current (NC) event sample at the near and far detectors adding background contributions due to misidentification of charged current as neutral current events. For this channel we assume a conservative 20\% energy resolution and 90\% detection efficiency, as was assumed for DUNE~\cite{Coloma:2017ptb}. Also this is a conservative choice, since better efficiencies have already been reached in Super-Kamiokande, which also uses water as detection material~\cite{T2K:2019zqh}. 
In former analyses the authors mostly considered a single detector simulation, assuming individual normalization and shape uncertainties in each channel. Since we perform a 2-detector analysis, we can correlate many of the systematic uncertainties among detectors and channels, improving the overall sensitivity of the experiment. The systematic uncertainties that we consider are essentially those of Ref.~\cite{Coloma:2012ji} for a superbeam experiment and are listed in Table~\ref{tab:sys_unc}. They include an overall uncertainty on the fiducial volume of the near and far detectors. This uncertainty is correlated between the neutrino and antineutrino modes. Next, there is an uncertainty on the main flux component, $\nu_\mu$ ($\overline{\nu}_\mu$) in neutrino (antineutrino) mode, and an individual uncertainty for each of the intrinsic background components. These uncertainties are uncorrelated between the neutrino and antineutrino modes, but must be correlated among the detectors and different channels of the same mode. We include a normalization uncertainty on each individual CC cross section for neutrinos and antineutrinos independently, but we also include penalties on the ratios between $\nu_\mu/\nu_e$ and $\overline{\nu}_\mu/\overline{\nu}_e$ cross sections, since they might be better constrained than individual measurements. Similarly, we include an uncertainty on the ratio of cross sections between neutrinos and antineutrinos. Since we include NC interactions in our analyses, there is an additional uncertainty on the NC cross section. In order to be conservative, in the NC channels we do not only use the ``Flux error background" from Table~\ref{tab:sys_unc}, due to flux-contamination, but we also include an additional uncertainty on the overall background. In our analyses we will consider an optimistic, a default, and a conservative set of uncertainties as indicated in Table~\ref{tab:sys_unc}. It should be noted that recently an analysis has been performed updating the estimates for some of the cross section related uncertainties~\cite{Dieminger:2023oin}. The authors obtained that most of the uncertainties for the ratios lie between our default and optimistic choices, but the uncertainty for the $\overline{\nu}_e/\overline{\nu}_\mu$ ratio is even better than our optimistic choice. Using these new values as default (and, e.g., the new values scaled by 1/2 and 2 factors for the optimistic and conservative choices) would only have marginal impact on our results. 

For our statistical analysis we use the $\chi^2$ function
\begin{equation}
    \chi^2(\vec{p})=\min_{\vec{\alpha}}\left\{ \sum_{\text{D}}\sum_{\text{C}} 2 \sum_{i=1}^{50} P^{\text{DC}}_i(\vec{p},\vec{\alpha})-D^{\text{DC}}_i+D^{\text{DC}}_i \log(D^{\text{DC}}_i/P^{\text{DC}}_i(\vec{p},\vec{\alpha})) + \sum_k (\alpha_k/\sigma_k)^2\right\}+\chi^2_\text{pen.}\,,
\end{equation}
where $\vec{p}$ is a set of neutrino oscillation parameters, $\vec{\alpha}$ are the systematic uncertainties with corresponding penalties $\sigma_k$, $D^{\text{DC}}_i$ is the simulated fake data in the $i$th energy bin for the detector D and the channel C, and $P^{\text{DC}}_i(\vec{p},\vec{\alpha})$ is the corresponding prediction in the same bin, which depends on the oscillation parameters $\vec{p}$ and the systematic uncertainties $\vec{\alpha}$. The first two sums are taken over the near and far detectors (D) and over the different detection channels (C). Finally,

\begin{equation}
\chi^2_\text{pen.} = \left(\frac{\theta_{13} - \theta^{\text{GF}}_{13}}{\sigma(\theta^{\text{GF}}_{13})}\right)^2
\label{eq:theta13_pen}
\end{equation}
is a penalty for $\theta_{13}$, where $\theta^{\text{GF}}_{13}\pm\sigma(\theta^{\text{GF}}_{13}) = (8.53 \pm 0.13)^{\circ}$ is taken from the global fit in Ref.~\cite{deSalas:2020pgw}.

\begin{table}[t]
  \begin{center}
    \begin{tabular}{|p{5cm}|ccc|}
      \hline
      Uncertainty & Opt. & Def. & Con.  \\
      \hline
      Fiducial volume ND                    & 0.2\% & 0.5\% & 1\%    \\
      Fiducial volume FD                    & 1\% & 2.5\% &  5\%     \\
      Flux error signal $\nu$               & 5\% & 7.5\% & 10\%     \\
      Flux error background $\nu$           & 10\% & 15\% & 20\%     \\
      Flux error signal $\bar\nu$           & 10\% & 15\% & 20\%     \\
      Flux error background $\bar\nu$       & 20\% & 30\% & 40\%     \\
      NC background                         & 5\% & 7.5\% & 10\%     \\ 
      CC cross section                      & 10\% & 15\% & 20\%     \\
      NC cross section                      & 10\% & 15\% & 20\%     \\
      $\nu_e$/$\nu_\mu$ ratio               & 3.5\% & 11\% & 22\%    \\
      $\nu$/$\overline\nu$ ratio            & 3.5\% & 11\% & 22\%    \\ 
      \hline
    \end{tabular}
  \end{center}

  \caption{\label{tab:sys_unc} The systematic uncertainties considered in this paper. See the text for more details.}
\end{table}

\section{CP-sensitivity}
\label{sec:SM}

In this section we compare the sensitivity to measure CP violation for our 2-detector configuration with and without NC channels with the single detector analysis. We calculate the sensitivity to exclude $\delta_{13} = 0$ and $\delta_{13} = \pi$ as a function of the true value of $\delta_{13}$. In order to create the fake data, we have fixed the remaining parameters to 
\begin{eqnarray}
    \Delta m_{21}^2 = 7.5\times10^{-5} \text{eV}^2, \qquad \Delta m_{31}^2 = 2.5\times10^{-3} \text{eV}^2\nonumber\\
    \sin^2\theta_{12}=0.32,\quad\sin^2\theta_{13}=0.022,\quad\sin^2\theta_{23}=0.5\,,
\end{eqnarray}
in accordance with the allowed regions obtained from recent global analyses~\cite{deSalas:2020pgw,Capozzi:2021fjo,Esteban:2020cvm}. In the analysis we marginalize over $\theta_{13}$ (using the prior from Ref.~\cite{deSalas:2020pgw}) and $\theta_{23}$ (freely). Since the upcoming JUNO experiment~\cite{JUNO:2022mxj} is going to measure soon the other three parameters with a precision below 1\%, we keep them fixed in the analysis. 

\begin{figure}
  \centering
    \includegraphics[width=0.49\textwidth]{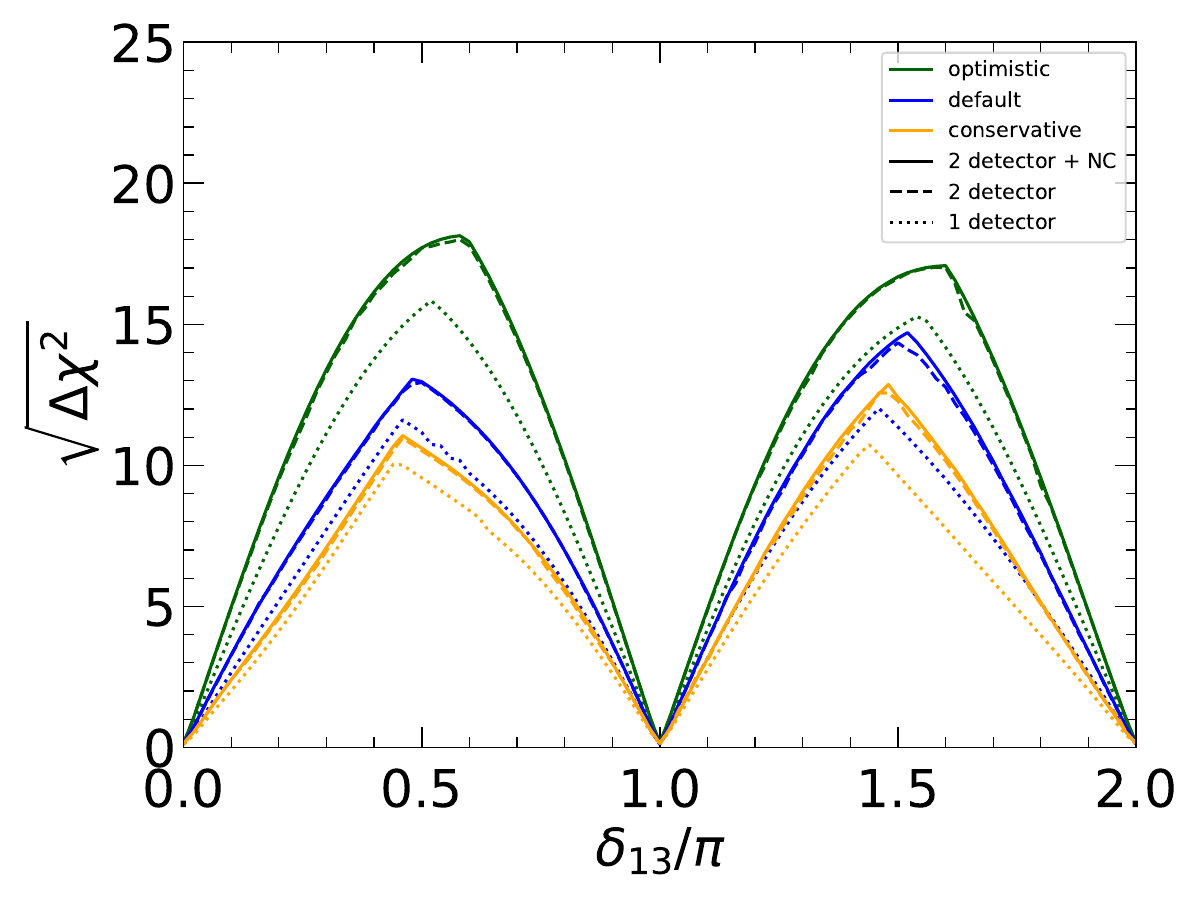}
    \caption{The sensitivity to measure CP violation at ESS$\nu$SB for different choices of systematic uncertainties and detector configurations.
    } 
  \label{fig:CP-3n}
\end{figure}

Figure~\ref{fig:CP-3n} shows the results of our analyses using only the far detector (dotted lines), using both detectors (dashed lines), and using both detectors including NC channels (solid lines) for the conservative (orange), default (blue) and optimistic (green) set of systematic uncertainties from Table~\ref{tab:sys_unc}. We find that the inclusion of the near detector improves the sensitivity, while the addition of NC channels has only a marginal effect. Note that the sensitivity in the default setting is very similar to the sensitivity presented by the ESS$\nu$SB collaboration in Ref.~\cite{ESSnuSB:2021azq}, even though the treatment of systematic uncertainties is much simpler there. The reason is that the simplified set of uncertainties used in Ref.~\cite{ESSnuSB:2021azq} represents sufficiently well the more realistic set of uncertainties considered here. The sensitivity could be further enhanced by adding an atmospheric neutrino sample, which could be detected at the far detector~\cite{Blennow:2019bvl}.  Another way to improve the sensitivity could be to make use of the better sensitivity of other experiments to measure $\theta_{23}$. We could further enhance the sensitivity if we imposed a prior on $\theta_{23}$ using the results of DUNE or T2HK in the same way we did for $\theta_{13}$ in Eq.~\eqref{eq:theta13_pen}. Since the beginning of data taking at ESS$\nu$SB is planned to start after DUNE or T2HK, it is most likely that there will be already results from these experiments on the $\theta_{23}$ measurement.

\section{Non-standard interactions}
\label{sec:NSI}

We turn our attention now to some BSM scenarios and discuss the sensitivity of ESS$\nu$SB to the parameters in each scenario.
In this section we consider neutral-current (NC) neutrino non-standard interactions (NSI)
generated by very heavy vector mediators
which are described by the effective four-fermion
interaction Lagrangian
(see the reviews in Refs.~\cite{Ohlsson:2012kf,Miranda:2015dra,Farzan:2017xzy,Proceedings:2019qno})
\begin{equation}
\mathcal{L}_{\text{NSI}}^{\text{NC}}
=
- \sqrt{2} G_{\text{F}}
\sum_{\alpha,\beta=e,\mu,\tau}
\overline{\nu_{\alpha L}} \gamma^{\rho} \nu_{\beta L}
\sum_{f=e,u,d}
\overline{f} \gamma_{\rho}
\left(
\varepsilon_{\alpha\beta}^{fV} - \varepsilon_{\alpha\beta}^{fA} \gamma_{5}
\right) f
,
\label{eq:NSI1}
\end{equation}
where $G_{\text{F}}$ is the Fermi constant.
This choice of parameterization of the NSI interaction Lagrangian
is useful because the Standard Model neutral-current weak interaction
Lagrangian can be obtained with the substitutions
$\varepsilon_{\alpha\beta}^{fV} \to g_V^f \delta_{\alpha\beta}$
and
$\varepsilon_{\alpha\beta}^{fA} \to g_A^f \delta_{\alpha\beta}$,
where $g_V^f$ and $g_A^f$ are the Standard Model
vector and axial neutral current couplings of the fermion $f$
(see, e.g., Table~3.6 of Ref.~\cite{Giunti:2007ry}).
Therefore,
the ratios
$\varepsilon_{\alpha\beta}^{fV}/g_V^f$
and
$\varepsilon_{\alpha\beta}^{fA}/g_A^f$
describe the size of non-standard interactions relative to
the Standard Model vector and axial neutral-current weak interactions.
Note that, contrary to the Standard Model neutral-current weak interactions,
non-standard interactions can generate neutrino flavor transitions
for $\alpha\neq\beta$.

The effective Hamiltonian which describes neutrino propagation in unpolarized matter
depends on the following combinations of the vector NSI couplings:
\begin{equation}
\varepsilon_{\alpha\beta}
=
\varepsilon_{\alpha\beta}^{eV}
+
\dfrac{N_{u}}{N_{e}}
\,
\varepsilon_{\alpha\beta}^{uV}
+
\dfrac{N_{d}}{N_{e}}
\,
\varepsilon_{\alpha\beta}^{dV}\,,
\label{eq:NSI2}
\end{equation}
where
$N_{e}$,
$N_{u}$, and
$N_{d}$
are, respectively, the number densities of
electrons, up quarks, and down quarks.
Since in long-baseline neutrino oscillation experiments
neutrinos propagate in the crust of the Earth where
the electron, proton and neutron number densities are approximately equal,
we consider the effective NSI parameters~\cite{Chatterjee:2020kkm,Denton:2020uda}
\begin{equation}
\varepsilon_{\alpha\beta}
=
\varepsilon_{\alpha\beta}^{eV}
+
3 \varepsilon_{\alpha\beta}^{uV}
+
3 \varepsilon_{\alpha\beta}^{dV}
.
\label{eq:NSI3}
\end{equation}
In general, these NSI parameters are complex and can be written as
\begin{equation}
\varepsilon_{\alpha\beta}
=
|\varepsilon_{\alpha\beta}| e^{i \phi_{\alpha\beta}}
.
\label{eq:NSI4}
\end{equation}
We take into account the constraint
$\varepsilon_{\alpha\beta} = \varepsilon_{\beta\alpha}^{*}$,
which follows from the hermiticity of the Lagrangian.

In Refs.~\cite{Chatterjee:2020kkm,Denton:2020uda},
these neutral-current non-standard interactions
have been suggested as a solution of the disagreement of the measurements of $\delta_{13}$ in the T2K~\cite{T2K:2021xwb} and NOvA~\cite{NOvA:2021nfi}
experiments.
Here we investigate if ESS$\nu$SB can test the region of parameter space preferred by these analyses. The solution of the disagreement of the measurements of $\delta_{13}$ in the T2K~\cite{T2K:2021xwb} and NOvA~\cite{NOvA:2021nfi}
experiments obtained in Refs.~\cite{Chatterjee:2020kkm,Denton:2020uda}
with non-standard interactions requires non-zero values of
$\epsilon_{e\mu}$ or $\epsilon_{e\tau}$.
In order to see if ESS$\nu$SB can test the allowed regions of these NSI
parameters obtained in Refs.~\cite{Chatterjee:2020kkm,Denton:2020uda},
we consider them one at a time,
as done in Refs.~\cite{Chatterjee:2020kkm,Denton:2020uda}.
In other words, when we consider $\epsilon_{e\mu}$ ($\epsilon_{e\tau}$)
all the other NSI parameters are kept fixed at $\epsilon_{\alpha\beta}=0$,
including $\epsilon_{e\tau}$ ($\epsilon_{e\mu}$).
On the other hand,
as in the Section~\ref{sec:SM},
we marginalize over the standard neutrino mixing parameters, 
which in this case include also $\delta_{13}$.

\begin{figure}
  \centering
    \includegraphics[width=0.49\textwidth]{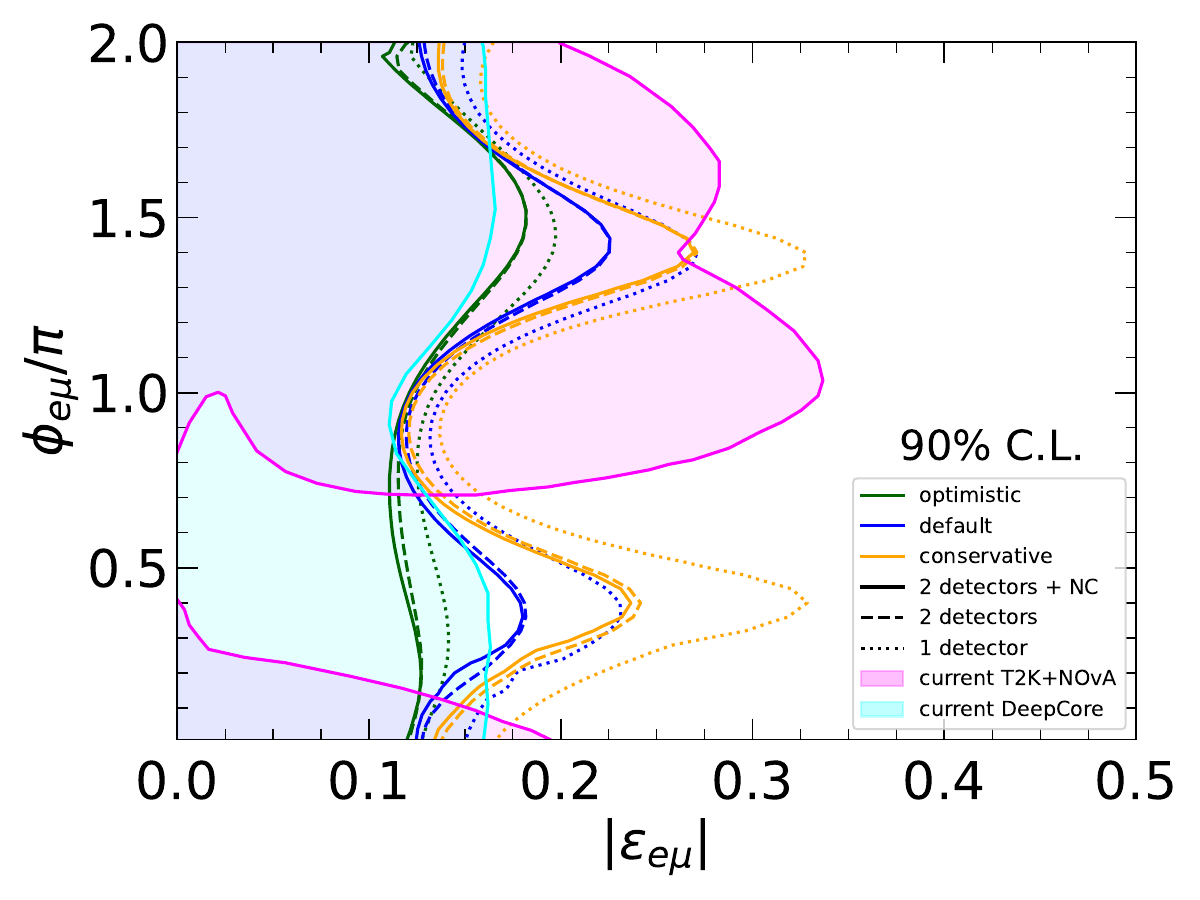}
    \includegraphics[width=0.49\textwidth]{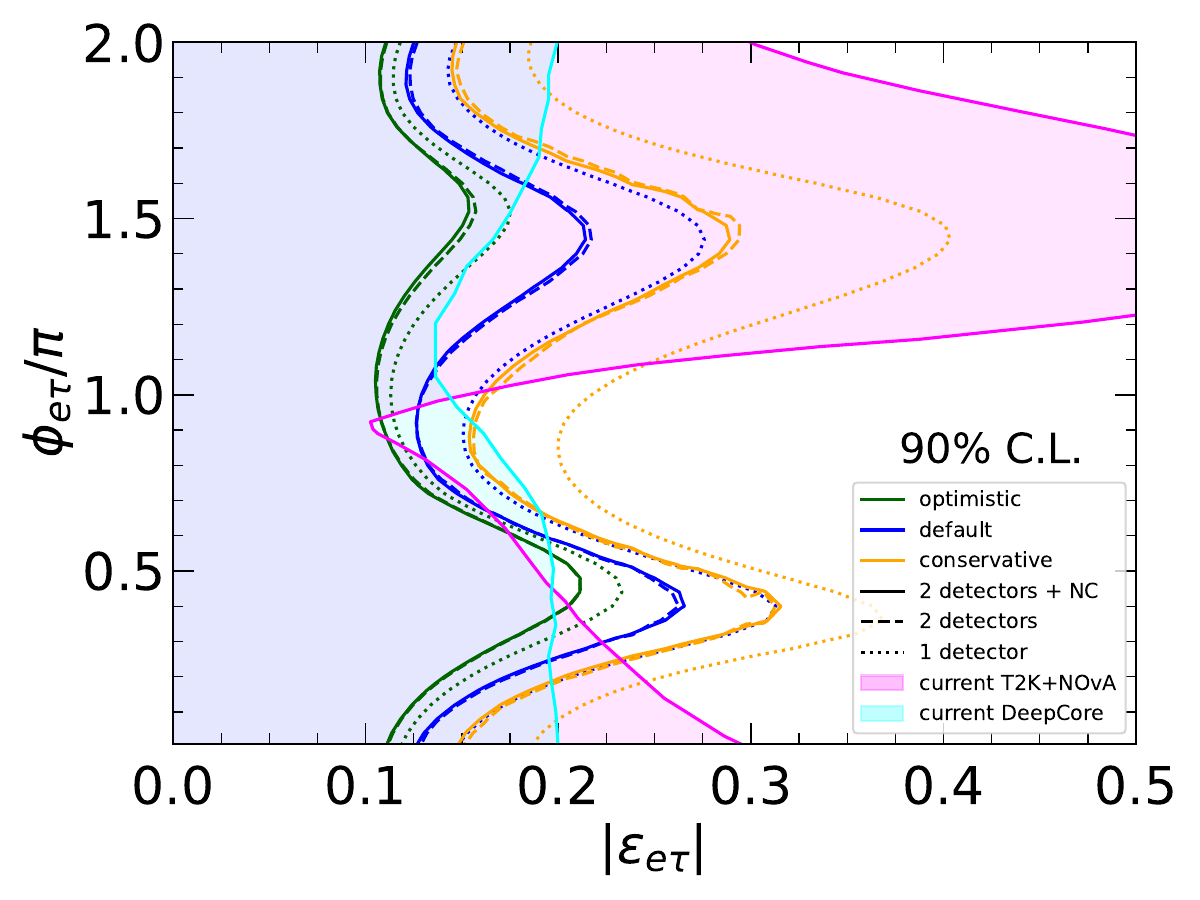}
    \caption{The expected exclusion region in the $\epsilon_{e\mu}$-$\delta_{e\mu}$ and  $\epsilon_{e\tau}$-$\delta_{e\tau}$ planes at 90\% C.L. for different choices of systematic uncertainties. Also shown are the allowed region extracted from Ref.~\cite{Chatterjee:2020kkm} and the bounds from Ref.~\cite{IceCube:NSI}.} 
  \label{fig:sens_NSI}
\end{figure}

The results of our analysis are presented in Fig.~\ref{fig:sens_NSI},
which shows the sensitivity regions of ESS$\nu$SB in the
$|\epsilon_{e\mu}|$-$\phi_{e\mu}$ and
and
$|\epsilon_{e\tau}|$-$\phi_{e\tau}$ planes at 90\% confidence level for the 1-detector (dotted) and 2-detector analysis without (dashed) and with (solid) NC channels for the optimistic (green), default (blue) and conservative (orange) choice of systematic uncertainties, as discussed in Section~\ref{sec:sim}.
In Fig.~\ref{fig:sens_NSI},
we also reproduce in magenta the boundary of the $90\%$ allowed regions extracted from Ref.~\cite{Chatterjee:2020kkm} and in cyan the one from the IceCube DeepCore analysis~\cite{IceCube:NSI}.

In the case of $\epsilon_{e\mu}$-$\phi_{e\mu}$ we find that ESS$\nu$SB will be able to disfavor an important part of the preferred region of Ref.~\cite{Chatterjee:2020kkm} at 90\% confidence. In the case of $\epsilon_{e\tau}$-$\phi_{e\tau}$ ESS$\nu$SB can exclude a larger part of the preferred region than in the $\epsilon_{e\mu}$-$\phi_{e\mu}$ case. In both cases the ESS$\nu$SB sensitivities are competitive with the current bounds from IceCube DeepCore. 

\begin{figure}
  \centering
    \includegraphics[width=0.49\textwidth]{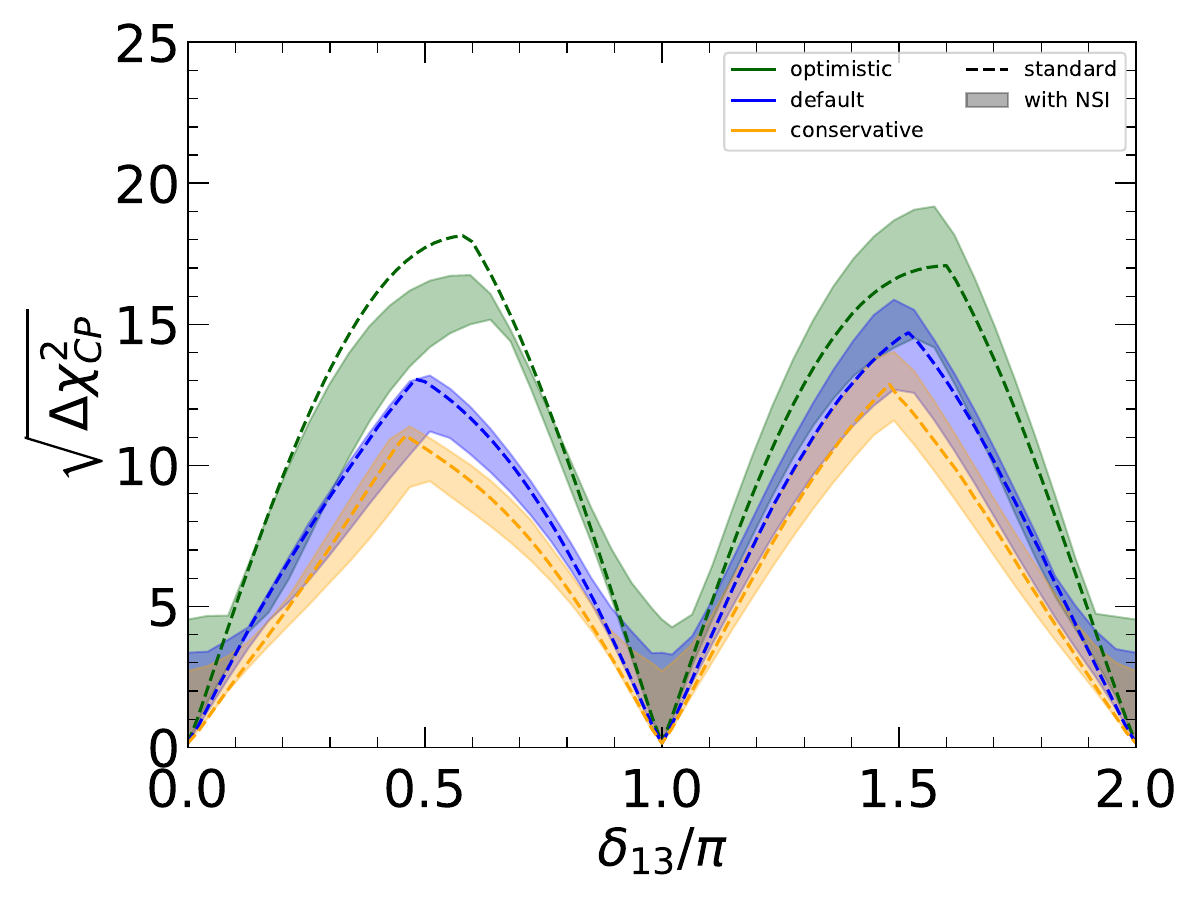}
    \caption{The impact NSI could have on the measurement of CP violation at ESS$\nu$SB for different choices of systematic uncertainties. The bands result from varying $\phi_{e\mu}$ in the fake data. The dashed lines, corresponding to the standard analysis without NSI, are the same as in Fig.~\ref{fig:CP-3n} for the analysis with two detectors and NC channels.
    } 
  \label{fig:CP-NSI}
\end{figure}

We also investigated the effect that NSI could have on the CP sensitivity of ESS$\nu$SB. To illustrate this problem we fixed $|\epsilon_{e\mu}| = 0.15$ (which is the best fit value of Ref.~\cite{Chatterjee:2020kkm}) and we calculated several fake data sets varying $|\phi_{e\mu}|$ and $\delta_{13}$. Next, we calculated the $\chi^2$ for excluding any CP conserving combination of CP phases. The results of this analysis are shown in Fig.~\ref{fig:CP-NSI}. The bands are generated by the variation of $\phi_{e\mu}$ in the fake data. It should be noted that these bands do not go to zero close to $\delta_{13}=0$ and  $\delta_{13}=\pi$. Indeed, for the case of optimistic systematic uncertainties they vary from 0 to about 4$\sigma$. This means that even if $\delta_{13}=0$ or $\delta_{13}=\pi$, ESS$\nu$SB can have the capability to measure possible CP violation due to $|\phi_{e\mu}|$ if $|\epsilon_{e\mu}|$ is large enough.

\section{Nonunitary neutrino mixing}
\label{sec:NU}

In this section we consider nonunitary neutrino mixing.
An effective nonunitary $3\times3$ neutrino mixing matrix can be the result of the presence of heavy neutral leptons which mix with the three light neutrinos,
but do not participate in neutrino oscillations
because they cannot be produced in low-energy experiments.
This can happen, for example, in the case of the Type-I seesaw mechanism.
In the traditional high-scale Type-I seesaw mechanism
the nonunitarity of the effective $3\times3$ neutrino mixing matrix
is negligible.
However, larger deviations from unitarity are generally expected
in low-scale Type-I seesaw models, such as those which implement the inverse and linear seesaw mechanisms~\cite{Mohapatra:1986bd,Akhmedov:1995ip,Akhmedov:1995vm,Malinsky:2005bi,Malinsky:2009gw,Malinsky:2009df,Hettmansperger:2011bt}.

If there are $n$ neutral leptons,
of which three are the ordinary light neutrinos
and $n-3$ are heavy neutral leptons,
the effective low-energy nonunitary $3\times3$ neutrino mixing matrix $N$
is a submatrix of the complete $n \times n$ unitary mixing matrix
\begin{equation}
 U^{n\times n} = 
 \begin{pmatrix}
  N & S
  \\
  V & T
 \end{pmatrix}\,,
\label{eq:Unxn}
\end{equation}
where $S$, $V$, and $T$ are, respectively,
$3\times(n-3)$,
$(n-3)\times3$, and
$(n-3)\times(n-3)$ matrices.
The nonunitary $3\times3$ mixing matrix $N$
can be written as~\cite{Escrihuela:2015wra}
\begin{equation}
 N = N^{NP}U=
 \begin{pmatrix}
  \alpha_{11} & 0 & 0
  \\
  \alpha_{21} & \alpha_{22} & 0
  \\
  \alpha_{31} & \alpha_{32} & \alpha_{33}
 \end{pmatrix}U\,,
\label{eq:N3x3}
\end{equation}
where $U$ is the standard unitary three-neutrino mixing matrix. The nonunitary new physics effects are encoded in the triangular matrix $N^{NP}$, which depends on  three real positive diagonal parameters $\alpha_{ii}$, and  three complex parameters $\alpha_{ij}$ for $i \neq j$.
Moreover,
the off-diagonal parameters are bounded by the inequality~\cite{Escrihuela:2016ube,Forero:2021azc}
\begin{equation}
 |\alpha_{ij}|
 \leq
 \sqrt{(1-\alpha_{ii}^2)(1-\alpha_{jj}^2)}\,,
 \label{eq:alphabound}
\end{equation}
and, for small unitarity violations,
the diagonal parameters are close to one.

The general expression of the probability of
$\nu_\alpha \to \nu_\beta$ oscillations in vacuum is
\begin{equation}
P_{\alpha\beta}
=
| ( N N^{\dagger} )_{\alpha\beta} |^2
-
4\sum_{k>j}\text{Re}\!\left[N_{\alpha k}^*N_{\beta k}N_{\alpha j}N_{\beta j}^*\right]\sin^2\!\left(\frac{\Delta m_{kj}^2L}{4E}\right)
+
2\sum_{k>j}\text{Im}\!\left[N_{\alpha k}^*N_{\beta k}N_{\alpha j}N_{\beta j}^*\right]\sin\!\left(\frac{\Delta m_{kj}^2L}{2E}\right)
.
\label{eq:nonuniprob}
\end{equation}
The first term of this probability is not equal to $\delta_{\alpha\beta}$ as in the unitary case and depends only on the values of the nonunitarity $\alpha$ parameters
(see Eq.~(A5) of Ref.~\cite{Forero:2021azc}).
Therefore, in the case of nonunitary neutrino mixing it is possible to
have a zero-distance flavor conversion.

In this section we discuss the sensitivity of ESS$\nu$SB to the nonunitarity parameters.
The role of the near detector is of particular importance,
because the zero-distance effect can alter the event rate at the near detector dramatically.

\begin{figure}
  \centering
    \includegraphics[width=0.49\textwidth]{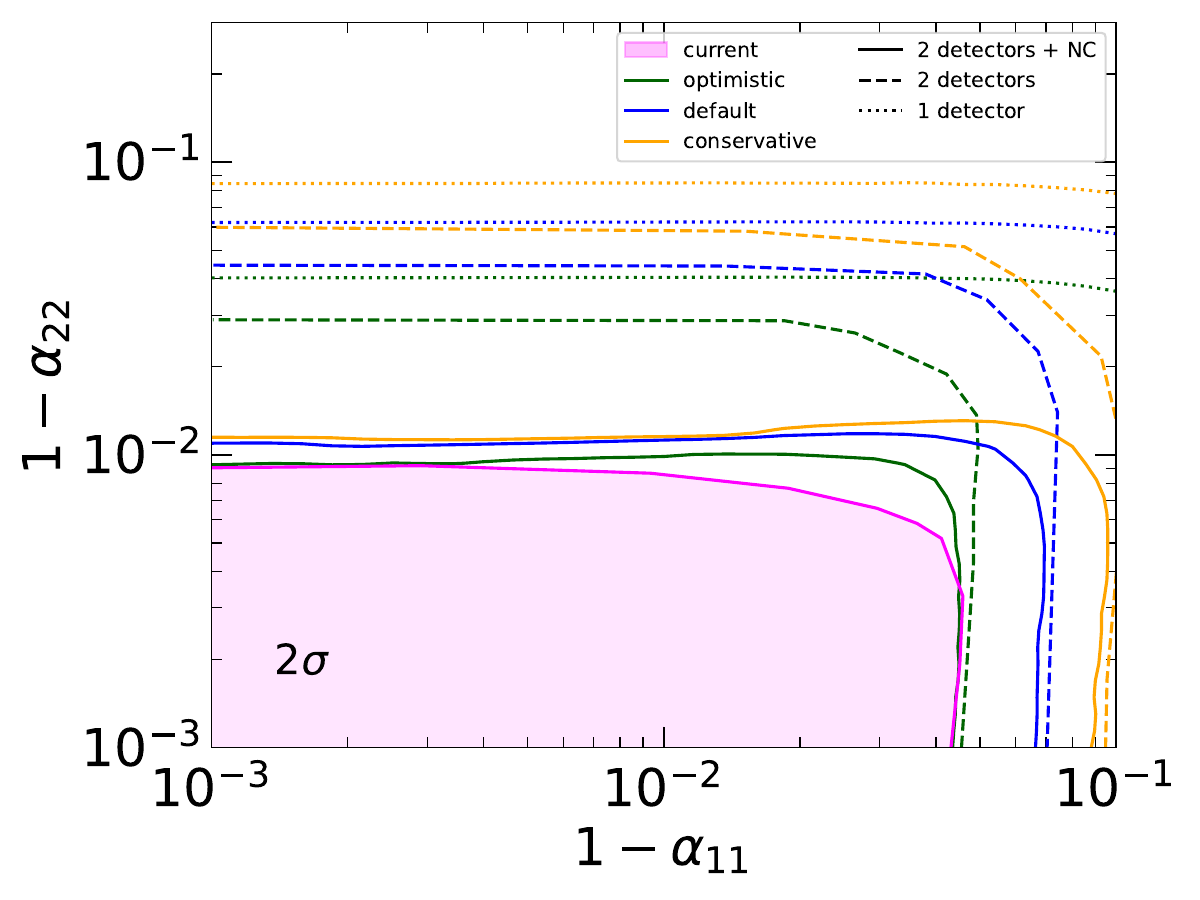}
    \includegraphics[width=0.49\textwidth]{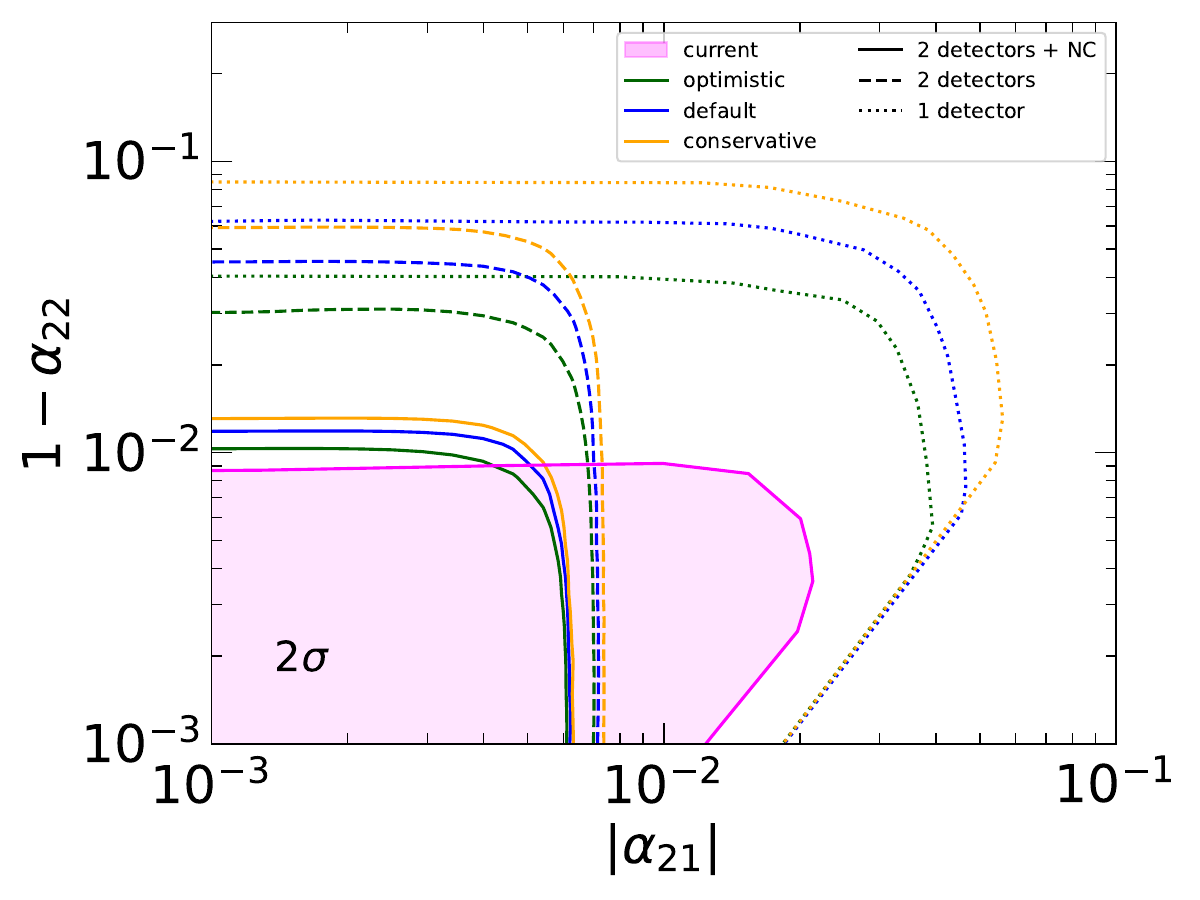}
    \caption{The expected region in different planes of the nonunitarity $\alpha$ parameters at 2$\sigma$ for different choices of systematic uncertainties. Also shown are the current bounds taken from Ref.~\cite{Forero:2021azc}.} 
  \label{fig:sens_NU}
\end{figure}

ESS$\nu$SB is mainly sensitive to the nonunitarity parameters
$\alpha_{22}$, 
$\alpha_{11}$, and
$|\alpha_{21}|$
through the $\nu_{\mu}$ and $\nu_{e}$ disappearance channels
and the $\nu_{\mu}\to\nu_{e}$ appearance channel.
The results of our analysis are shown in Fig.~\ref{fig:sens_NU},
which has been obtained by marginalizing
over the standard three neutrino mixing parameters,
over the phase of $\alpha_{21}$,
and over $|\alpha_{21}|$ ($\alpha_{11}$) in the left (right) panel.
Also shown, for the comparison with the sensitivity of ESS$\nu$SB, are the current bounds obtained in Ref.~\cite{Forero:2021azc}. We show again the ESS$\nu$SB sensitivities for the different choices of uncertainties and for the 1-detector and the 2-detector analyses with and without NC channels. As one can see from the figure, ESS$\nu$SB can only set bounds similar to the current ones on the diagonal parameters $\alpha_{11}$ and $\alpha_{22}$. However, an important improvement can be expected in the case of $|\alpha_{21}|$. It should also be noted that the inclusion of NC events helps to improve the sensitivity on each parameter importantly. 

\begin{figure}
  \centering
    \includegraphics[width=0.49\textwidth]{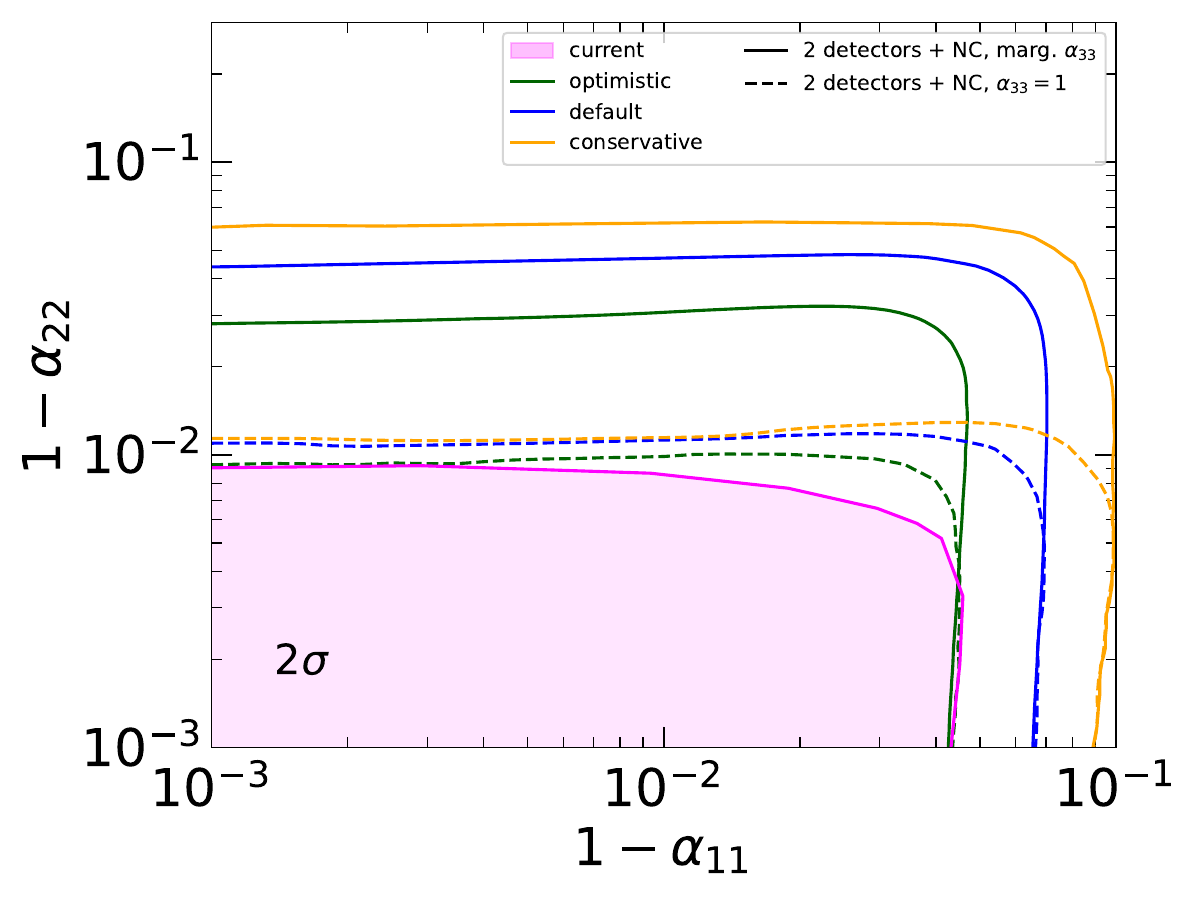}
    \includegraphics[width=0.49\textwidth]{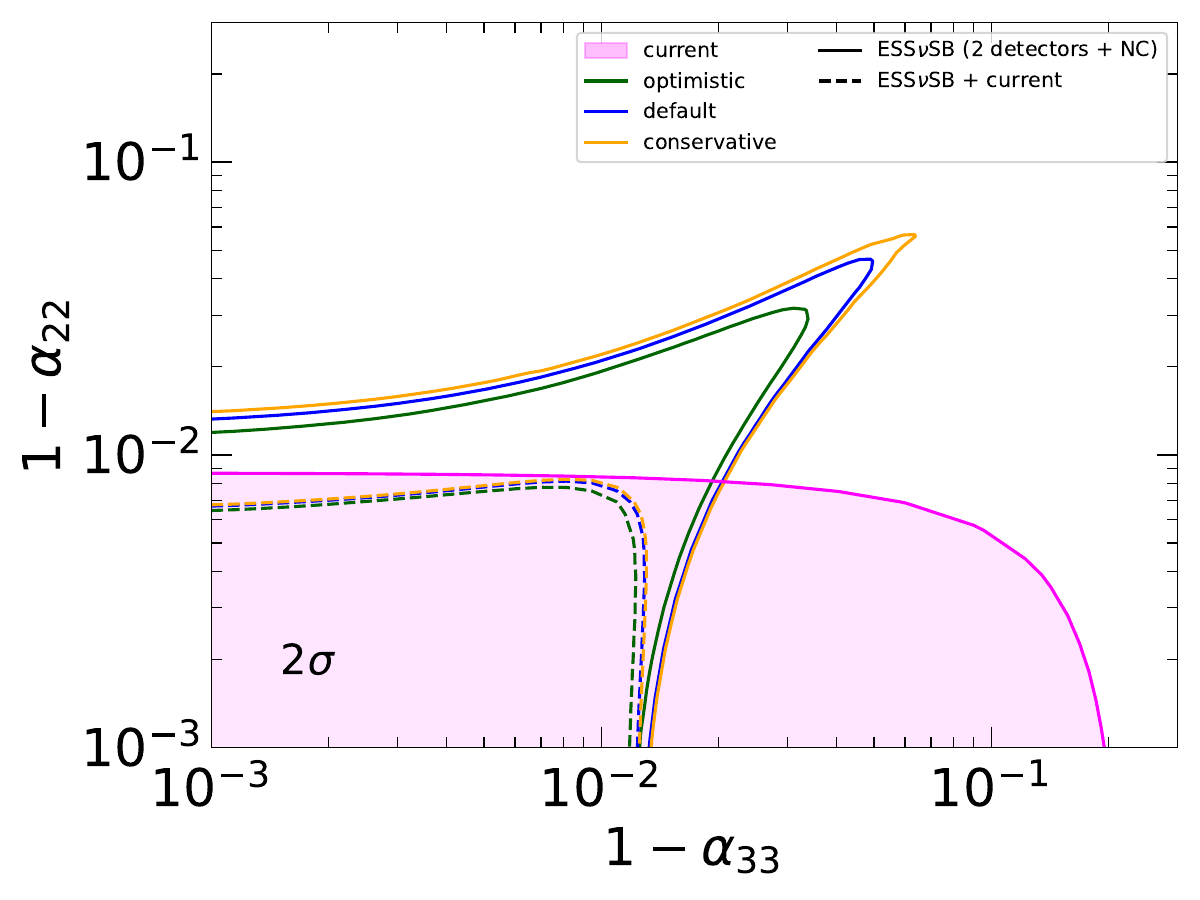}
    \caption{Left: Comparison of the ESS$\nu$SB sensitivities in the $(1-\alpha_{11})-(1-\alpha_{22})$ plane obtained marginalizing over $\alpha_{33}$ (solid lines) and keeping $\alpha_{33}=1$ fixed (dashed lines). The dashed lines in this figures are the same as the solid lines in Fig.~\ref{fig:sens_NU}. Right: The bounds in the $(1-\alpha_{33})-(1-\alpha_{22})$ plane than can be obtained from an analysis of ESS$\nu$SB data alone and from a combined analysis with the current bounds.
    In both panels we also show the current bounds taken from Ref.~\cite{Forero:2021azc}.} 
  \label{fig:sens_NU_marg3}
\end{figure}

The contours of Fig.~\ref{fig:sens_NU} have been obtained keeping $\alpha_{3j}$ fixed at their unitary values
($\alpha_{31}=\alpha_{32}=0$ and $\alpha_{33}=1$). However, they can affect the ESS$\nu$SB sensitivity since they change the oscillation probability through matter effects. On the other hand, we can also bound these parameters directly by observation of the NC channels\footnote{Since the ESS$\nu$SB sensitivity to the $\alpha_{3j}$ parameters without NC channels is very weak, we do not discuss it here.}. This is possible because the NC sample can be used to measure
\begin{equation}
  P_\alpha^{\text{NC}} = 1 - P_{\alpha s} = \sum_{\beta = e,\mu,\tau} P_{\alpha\beta}\,,
\end{equation}
where $\alpha=e,\mu$ and $P_{\alpha s}$ means the oscillation of an initial flavor $\alpha$ into a sterile state $s$. This quantity depends on all parameters and therefore allows us to bound $\alpha_{3j}$.
From here, we focus the discussion on the analysis using both detectors and including the NC channels. In the left panel of Fig.~\ref{fig:sens_NU_marg3} we show how the marginalization of $\alpha_{33}$ (the effects of $|\alpha_{31}|$ and $|\alpha_{32}|$ are negligible in comparison with $\alpha_{33}$) affects the sensitivity. The solid lines have been obtained by marginalizing over $\alpha_{33}$,
while the dashed lines correspond to the previous analysis where $\alpha_{33}$ was kept fixed at the unitary value. As one can see, the sensitivity to $\alpha_{11}$ is practically unaffected by the different assumptions on the values of $\alpha_{33}$. On the other hand, the sensitivity to $\alpha_{22}$ is weakened by the marginalization over $\alpha_{33}$. We have also found that the marginalization over the $\alpha_{3j}$ parameters does not affect the sensitivity to $|\alpha_{21}|$. When bounding the non-diagonal parameters, two effects must be considered. First, the parameters can enter directly the oscillation probabilities. Second, if we obtain strong bounds on two diagonal parameters, we can place a strong bound on the corresponding non-diagonal parameter through Eq.~\eqref{eq:alphabound}. In our analysis the contribution of Eq.~\eqref{eq:alphabound} to the bound on $|\alpha_{21}|$ is weak. Therefore, the weaker bound on $\alpha_{22}$ that is obtained after marginalization over $\alpha_{33}$ does not affect the bound on $|\alpha_{21}|$.

In the right panel of Fig.~\ref{fig:sens_NU_marg3} we show the sensitivity in the $(1-\alpha_{33})-(1-\alpha_{22})$ plane. The solid lines represent the sensitivity of ESS$\nu$SB alone.
As can be seen, a correlation between the parameters reduces the sensitivity to $\alpha_{22}$ when $1-\alpha_{33}$ is relatively large. Note, that ESS$\nu$SB can improve the current bound (magenta lines) on $\alpha_{33}$. In Fig.~\ref{fig:sens_NU_marg3}, we also show the sensitivity that can be obtained by combining the ESS$\nu$SB sensitivity with the current bounds. Since the current bound on $\alpha_{22}$ is better than the ESS$\nu$SB sensitivity, the inclusion of the current bound breaks the degeneracy between the parameters, leading to a great improvement of the combined sensitivity to $\alpha_{33}$. As can be seen from the figure, the combined sensitivity to $\alpha_{33}$ marginalized over $\alpha_{22}$ is about one order of magnitude better than that of ESS$\nu$SB alone for any choice of systematic uncertainties.

\begin{figure}
  \centering
    \includegraphics[width=0.32\textwidth]{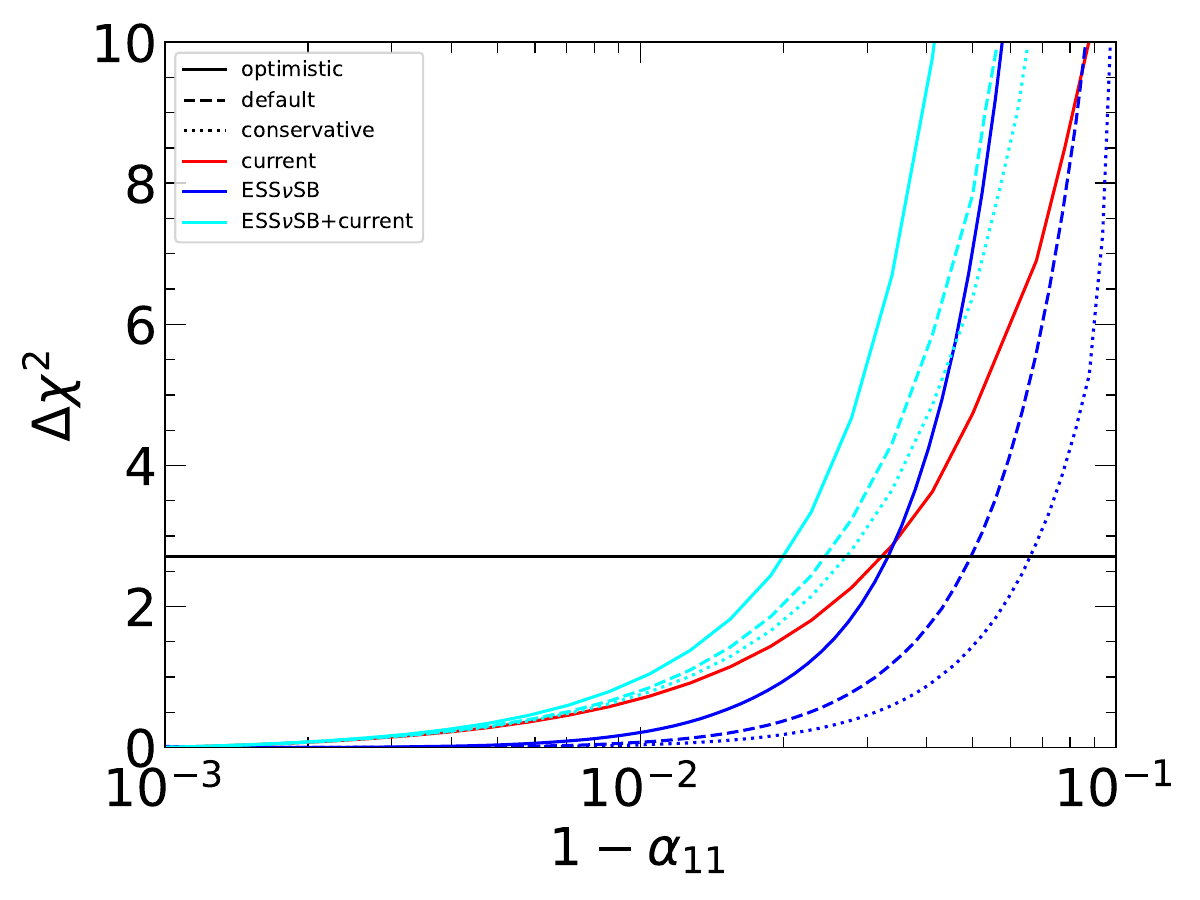}
    \includegraphics[width=0.32\textwidth]{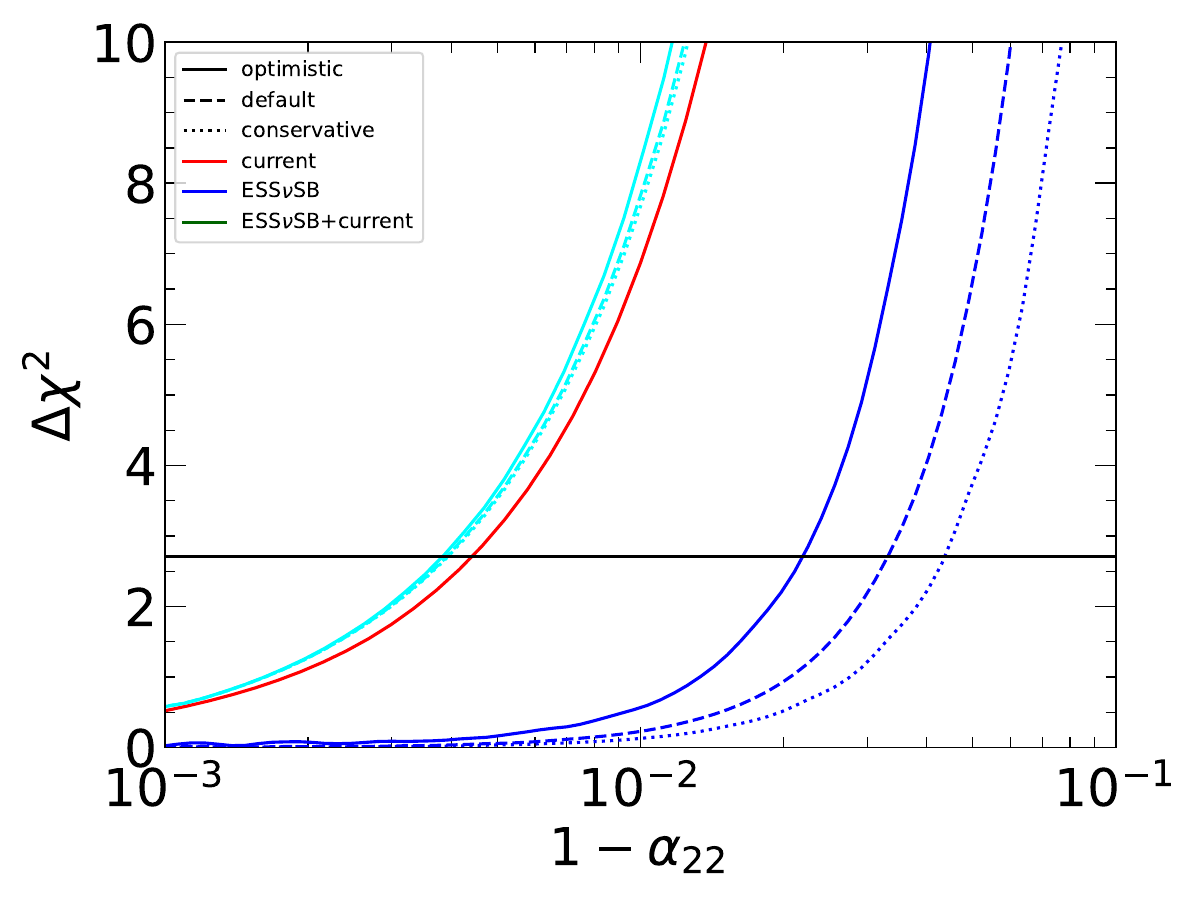}
    \includegraphics[width=0.32\textwidth]{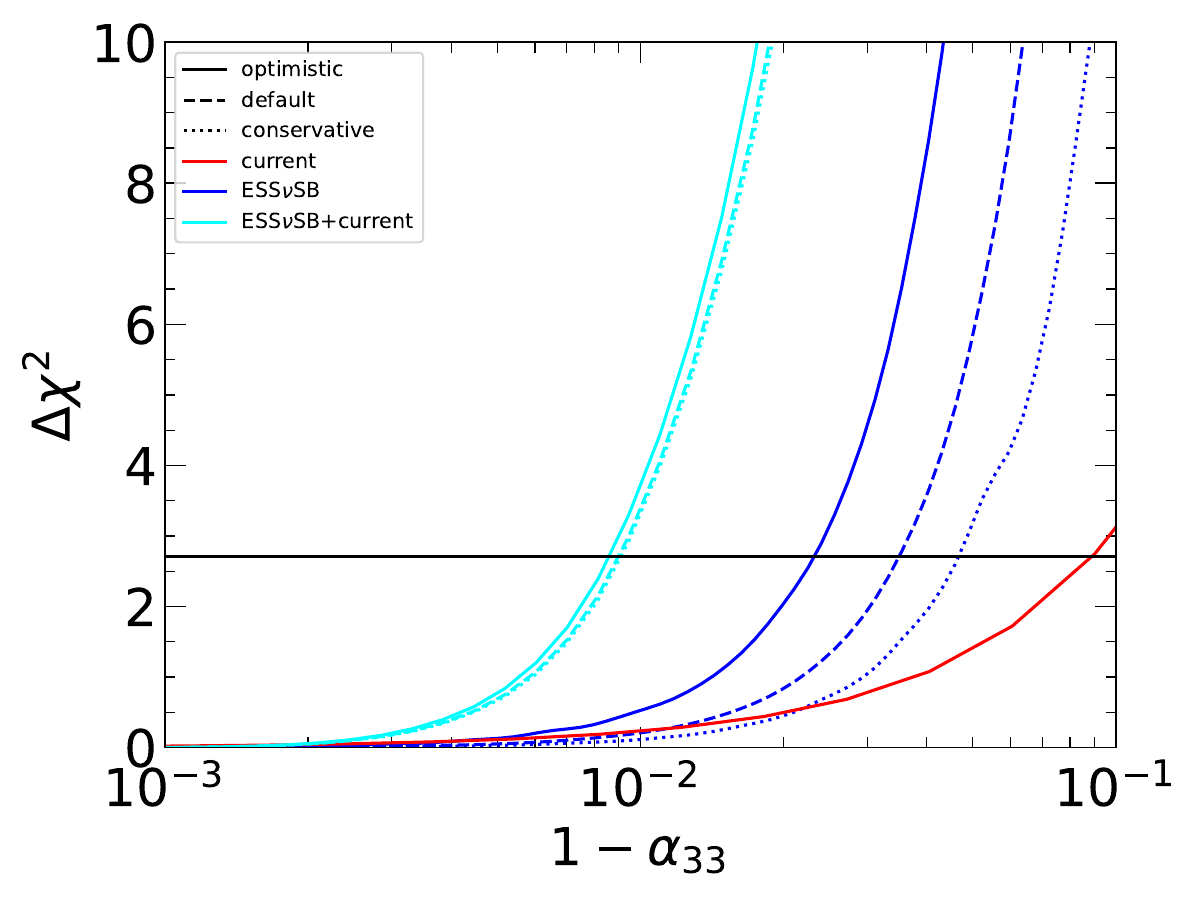}\\
    \includegraphics[width=0.32\textwidth]{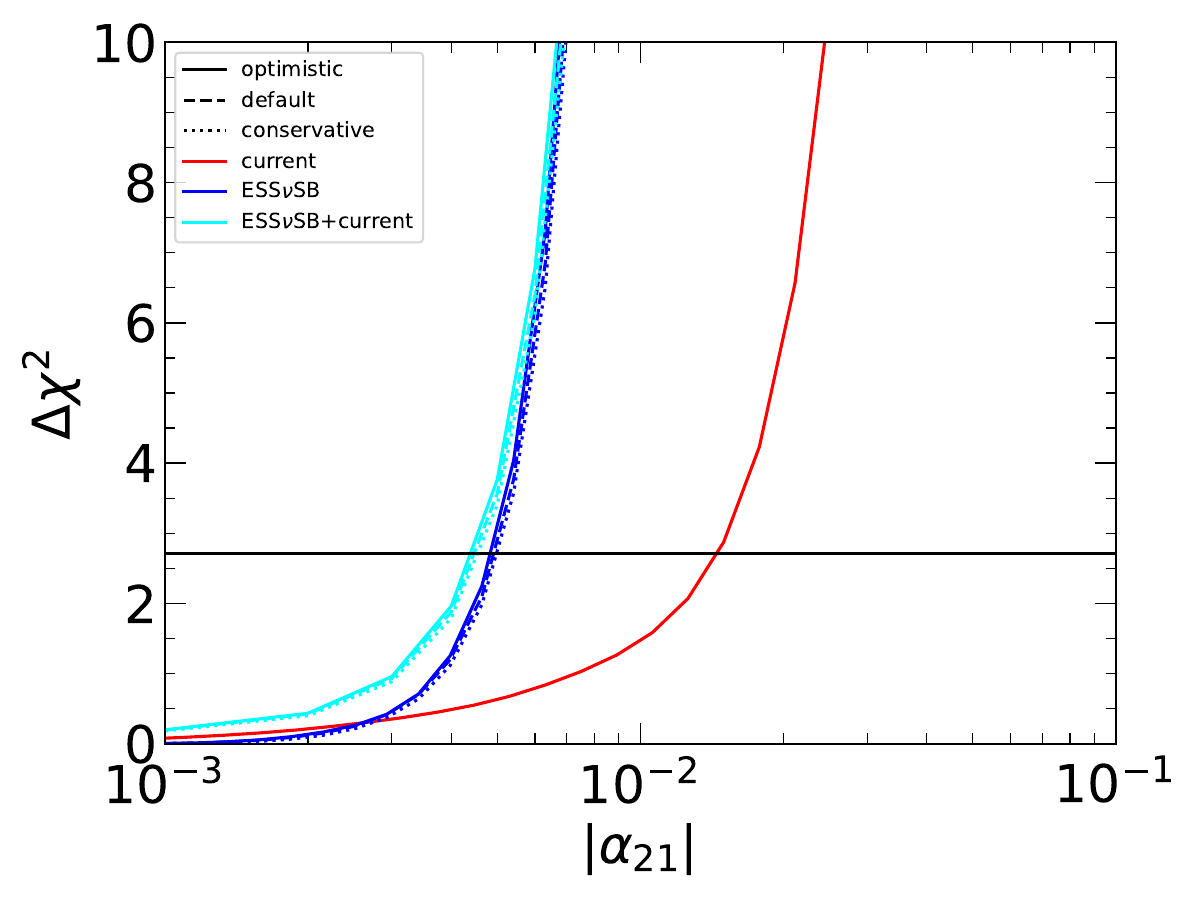}
    \includegraphics[width=0.32\textwidth]{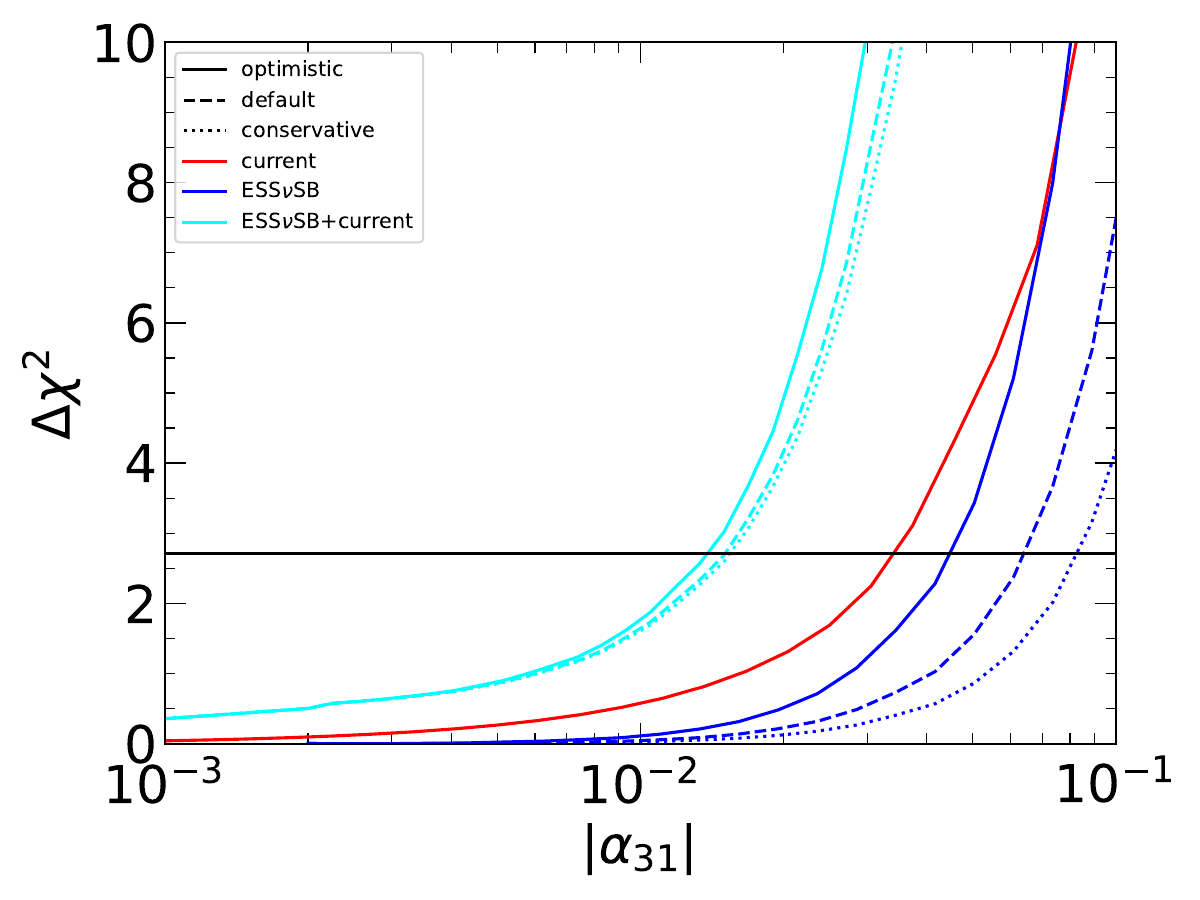}
    \includegraphics[width=0.32\textwidth]{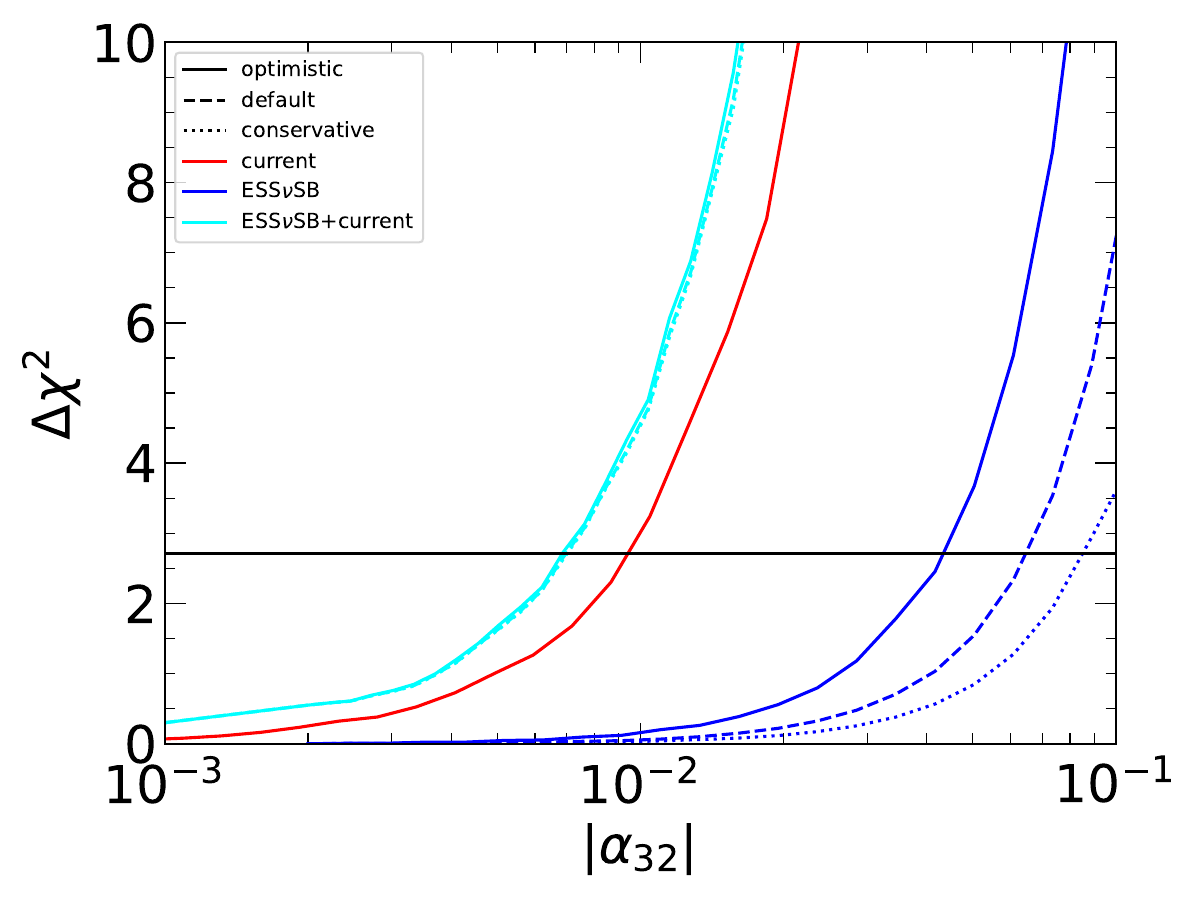}
    \caption{The $\Delta\chi^2$ profiles for the ESS$\nu$SB sensitivity for the 2-detector analysis including NC channels to the nonunitarity parameters (blue), in comparison with the profiles obtained in Ref.~\cite{Forero:2021azc} (red) and the profiles that could be obtained from a combined analysis of current data and ESS$\nu$SB (cyan) for different assumptions on the systematic uncertainties. The horizontal solid black line shows the $\Delta\chi^2=2.71$ corresponding to
90\% C.L.} 
  \label{fig:sens_NU_1D}
\end{figure}

In Fig.~\ref{fig:sens_NU_1D} we show the
sensitivity of ESS$\nu$SB to each of the nonunitarity parameters for the 2-detector analysis with NC channels for different choices of systematic uncertainties (blue lines) in comparison with the current bounds obtained in Ref.~\cite{Forero:2021azc} (red lines). We also show the results that could be obtained from a combined analysis using current data and ESS$\nu$SB (cyan lines).
The $\Delta\chi^2$ for each parameter has been obtained by marginalizing
over all the other parameters. This figure highlights the important improvement that can be expected on the bound on $|\alpha_{21}|$, which is only slightly affected by the choice of uncertainties, unlike in the case of $\alpha_{11}$ and $\alpha_{22}$, where the bounds get considerably worse when allowing for larger uncertainties. The lower panels show the bounds that can be obtained on the $\alpha_{3j}$ parameters. As can be seen in the figure, ESS$\nu$SB can improve the current bound on $\alpha_{33}$, but not on the non-diagonal $\alpha_{3j}$ parameters. However, once we combine with current data, the constraints on all parameters improve and become stronger than the current bounds. The improvement in $\alpha_{3j}$ is particularly strong. This happens due to the fact that, unlike in the case of $|\alpha_{21}|$, the main contribution to the bounds of $|\alpha_{31}|$ and $|\alpha_{32}|$ comes from Eq.~\eqref{eq:alphabound}. Therefore, an improvement of the bounds on $\alpha_{33}$ and $\alpha_{22}$ has an important impact on the sensitivity.

The sensitivity of ESS$\nu$SB to the nonunitarity of the
neutrino mixing matrix has been already studied
in Ref.~\cite{Chatterjee:2021xyu} considering only the far detector
in a one-detector analysis. The authors obtain similar bounds to ours in the case of $|\alpha_{21}|$. However, the bounds obtained for $\alpha_{11}$ and $\alpha_{22}$ are much stronger there. This happens due to a different treatment of systematic uncertainties. In Ref.~\cite{Chatterjee:2021xyu} only an overall uncertainty on the signal and background of each channel has been considered, while we treat each background component individually. Therefore, our more conservative treatment of uncertainties leads to more conservative sensitivities.

\begin{figure}
  \centering
    \includegraphics[width=0.49\textwidth]{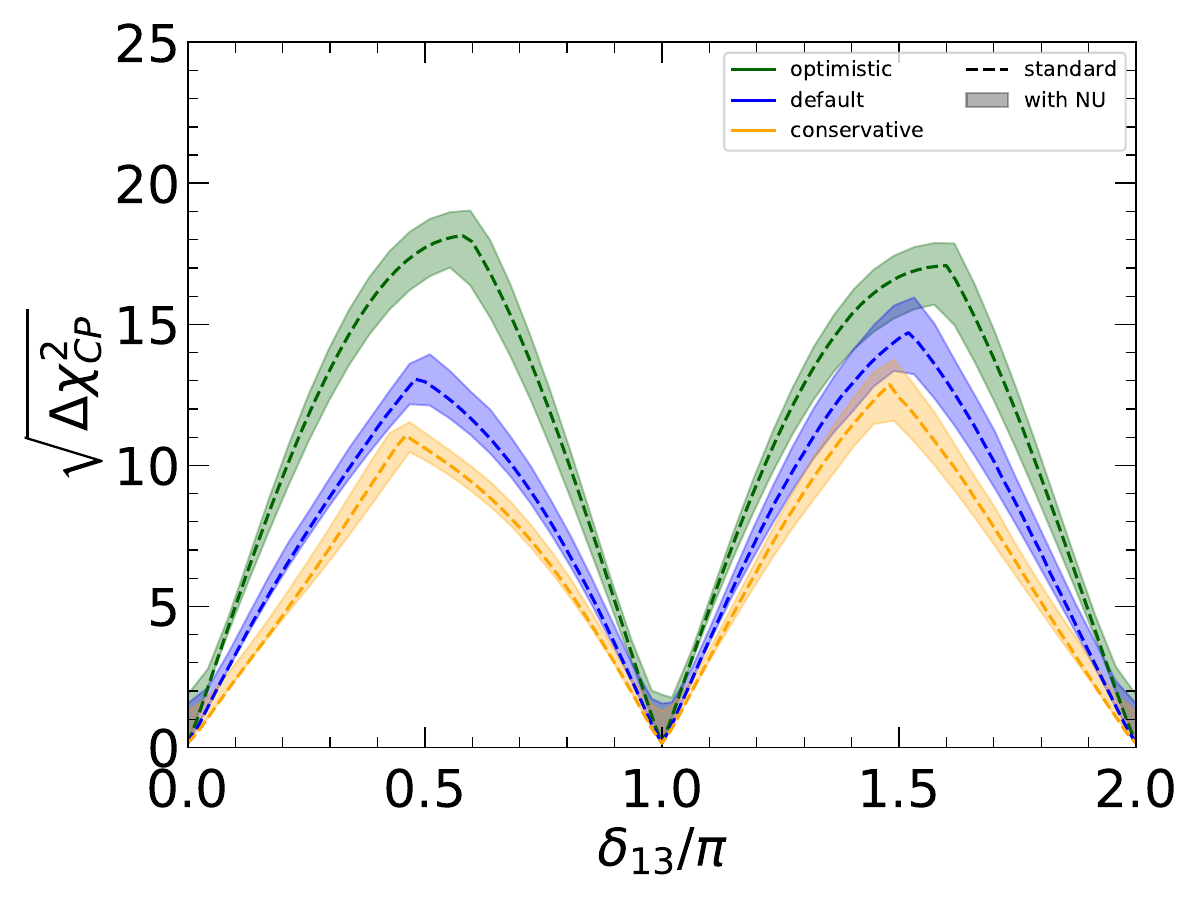}
    \caption{The impact nonunitary neutrino mixing could have on the measurement of CP violation at ESS$\nu$SB for different choices of systematic uncertainties. The width of the bands result from varying $\phi_{21}$ in the fake data. The dashed lines correspond to the standard analyses and are the same as in Fig.~\ref{fig:CP-3n}.} 
  \label{fig:CP_NU}
\end{figure}

Finally, let us discuss how nonunitary neutrino mixing could affect the CP sensitivity at ESS$\nu$SB. In order to show the impact, we fixed $\alpha_{11} = 0.99$, $\alpha_{22} = 0.998$, and $|\alpha_{21}|=0.007$. These values lie within the bounds of the current analysis in Ref.~\cite{Forero:2021azc}, while lying within the sensitivity range of ESS$\nu$SB (for $|\alpha_{21}|$). We generated fake data sets varying $\delta_{13}$ and $\phi_{21}$, which is the CP phase associated to $|\alpha_{21}|$, and we computed the sensitivity to exclude all CP-conserving combinations of phases. We consider only the case of two detectors with NC channels here. The results of this analysis are shown in Fig.~\ref{fig:CP_NU}. The value of $\delta_{13}$ is varied on the $x$-axis and the width of the bands is obtained from varying $\phi_{21}$. The dashed lines, corresponding to the standard analysis, are the same as those obtained in Sec.~\ref{sec:SM}. As can be seen, the presence of nonunitarity has small effects on the overall CP sensitivity. It should be noted that the bands do not go to zero around $\delta_{13}=0$ (and $2\pi$) and $\delta_{13}=\pi$. This is due to the fact that even if there is no CP violation due to $\delta_{13}$, there might be CP violation due to $\phi_{21}$ to which ESS$\nu$SB is sensitive. Here we have discussed only the effect of $\phi_{21}$. The width of the bands could be slightly increased further if we added the contributions of $\phi_{31}$ and $\phi_{32}$. However, the effects of these phases are expected to be smaller than those of $\phi_{21}$. We remind that the bounds on $|\alpha_{31}|$ and $|\alpha_{32}|$ presented above have been obtained from the relation in Eq.~\eqref{eq:alphabound}, since these parameters have only marginal effect on the oscillation probabilities considered here. This means that if CP was violated due to $\phi_{31}$ or $\phi_{32}$ (and not $\phi_{21}$ or $\delta_{13}$), ESS$\nu$SB would most likely not see it.

\section{Light sterile neutrinos}
\label{sec:3+1}

In this section we study the ESS$\nu$SB sensitivity to neutrino oscillations
generated by the mixing of the three active neutrinos
$\nu_{e}$,
$\nu_{\mu}$,
$\nu_{\tau}$
with a sterile neutrino $\nu_{s}$
which is mainly composed by a new neutrino mass eigenstate
$\nu_{4}$ having a mass $m_{4} \gtrsim 1 \, \text{eV}$.
This mass is light,
but heavier than the masses
$m_{1}$,
$m_{2}$,
$m_{3}$
of the three standard neutrino mass eigenstates
$\nu_{1}$,
$\nu_{2}$,
$\nu_{3}$
which are constrained below the eV scale
by $\beta$ decay~\cite{KATRIN:2021uub},
neutrinoless double-$\beta$ decay~\cite{Agostini:2022zub}
and
cosmological~\cite{Planck:2018vyg} bounds.
In this scenario, called ``3+1'',
there is a new squared-mass difference
$\Delta{m}^2_{41} \equiv m_{4}^2 - m_{1}^2 \gtrsim 0.1~\text{eV}^2$
which is much larger than the solar and atmospheric neutrino oscillations
squared-mass differences, which generate the oscillations observed in
solar, atmospheric and long-baseline neutrino oscillation experiments
(see, e.g., the review in Ref.~\cite{ParticleDataGroup:2022pth}
and the recent three-neutrino global analyses in
Refs.~\cite{deSalas:2020pgw,Esteban:2020cvm,Capozzi:2021fjo}).
The new squared-mass difference
$\Delta{m}^2_{41}$
generates short-baseline neutrino oscillations which may explain,
at least partially,
the anomalies found in
short-baseline neutrino oscillation experiments:
the Gallium Anomaly,
the Reactor Antineutrino Anomaly,
and the LSND and MiniBooNE anomalies
(see the reviews in Refs.~\cite{Gariazzo:2015rra,Gonzalez-Garcia:2015qrr,Giunti:2019aiy,Diaz:2019fwt,Boser:2019rta,Dasgupta:2021ies,Acero:2022wqg}).

In the 3+1 scenario,
the effective probabilities of neutrino oscillations in vacuum
in short-baseline experiments are given by
\begin{equation}
P_{\nu_{\alpha}\to\nu_{\beta}}^{\text{SBL}}
=
\left|
\delta_{\alpha\beta}
-
\sin^2 2\vartheta_{\alpha\beta}
\sin^{2}\!\left( \frac{\Delta{m}^2_{41}L}{4E} \right)
\right|
,
\label{eq:probSBL}
\end{equation}
with the oscillation amplitudes given by the effective mixing parameters
\begin{equation}
\sin^2 2\vartheta_{\alpha\beta}
=
4
|U_{\alpha 4}|^2
\left| \delta_{\alpha\beta} -  |U_{\beta 4}|^2 \right|
.
\label{eq:ampSBL}
\end{equation}
These oscillation amplitudes depend on the absolute values of
the elements in the fourth column of the $4\times4$ unitary mixing matrix $U$.
Therefore, the effective
oscillation probabilities of neutrinos and antineutrinos
in short-baseline experiments
are equal.
There is, however, a difference between the oscillation probabilities of neutrinos and antineutrinos
at longer distances,
as those of long-baseline experiments~\cite{Klop:2014ima,Berryman:2015nua,Gandhi:2015xza,Palazzo:2015gja,Dutta:2016glq,Capozzi:2016vac,Fiza:2021gvq,Giarnetti:2021wur,Ghosh:2019zvl},
where the effects of all the complex phases in the mixing matrix
are observable.

The elements of the fourth column of the mixing matrix
must be small, because 3+1 active-sterile neutrino mixing
must be a small perturbation of the standard three-neutrino mixing
which fits very well the robust data of solar, atmospheric and long-baseline neutrino oscillation experiments~\cite{deSalas:2020pgw,Esteban:2020cvm,Capozzi:2021fjo}.

In the standard parameterization of the $4\times4$ unitary mixing matrix $U$
(see, e.g., Refs.~\cite{Giunti:2019aiy,Boser:2019rta})
we have
\begin{equation}
U_{e4}
=
\sin\vartheta_{14} \, e^{-i\delta_{14}}
\quad
\text{and}
\quad
U_{\mu4}
=
\cos\vartheta_{14}
\sin\vartheta_{24}
\simeq
\sin\vartheta_{24}
.
\label{eq:U4x4}
\end{equation}
The approximation takes into account the smallness of
the new mixing angle $\vartheta_{14}$.
Hence,
the effective mixing parameters in short-baseline
$\nu_{e}$ and $\nu_{\mu}$
disappearance experiments are given by, respectively,
\begin{align}
\null & \null
\sin^2 2\vartheta_{ee}
=
4 |U_{e4}|^2 \left( 1 -  |U_{e4}|^2 \right)
=
\sin^2 2\vartheta_{14}
,
\label{eq:see}
\\
\null & \null
\sin^2 2\vartheta_{\mu\mu}
=
4 |U_{\mu4}|^2 \left( 1 -  |U_{\mu4}|^2 \right)
\simeq
\sin^2 2\vartheta_{24}
.
\label{eq:smm}
\end{align}
The effective mixing parameter in short-baseline
$\nu_{\mu}\to\nu_{e}$
appearance experiments has the more complicated expression
\begin{equation}
\sin^2 2\vartheta_{\mu e}
=
4 |U_{e4}|^2 |U_{\mu4}|^2
=
\sin^2 2\vartheta_{14}
\sin^2\vartheta_{24}
\simeq
\dfrac{1}{4}
\,
\sin^2 2\vartheta_{ee}
\sin^2 2\vartheta_{\mu\mu}
.
\label{eq:sme}
\end{equation}
The approximation follows from the smallness of
the new mixing angles $\vartheta_{14}$ and $\vartheta_{24}$.

Considering the ESS$\nu$SB average neutrino energy
$E \simeq 400 \, \text{MeV}$~\cite{Alekou:2022mav},
the oscillation phase at the near detector ($L \simeq 250 \, \text{m}$) is
$\Delta{m}^2_{41}L/4E \gtrsim 0.8$
for $\Delta{m}^2_{41} \gtrsim 1 \, \text{eV}^2$.
Therefore,
the short-baseline oscillations generated by $\Delta{m}^2_{41}$
may be observable at the ESS$\nu$SB near detector
and average out at longer distances as that of the far detector.
Hence,
in a two-detector analysis the sensitivity
to active-sterile neutrino mixing is due to the ESS$\nu$SB near detector
and the far detector reduces the systematic uncertainties.  Note that in the calculation of the oscillation probability at the far detector matter effects must be included.

\begin{figure}
  \centering
\subfigure[]{ \label{fig:sq14_dm41}
\includegraphics[width=0.48\textwidth]{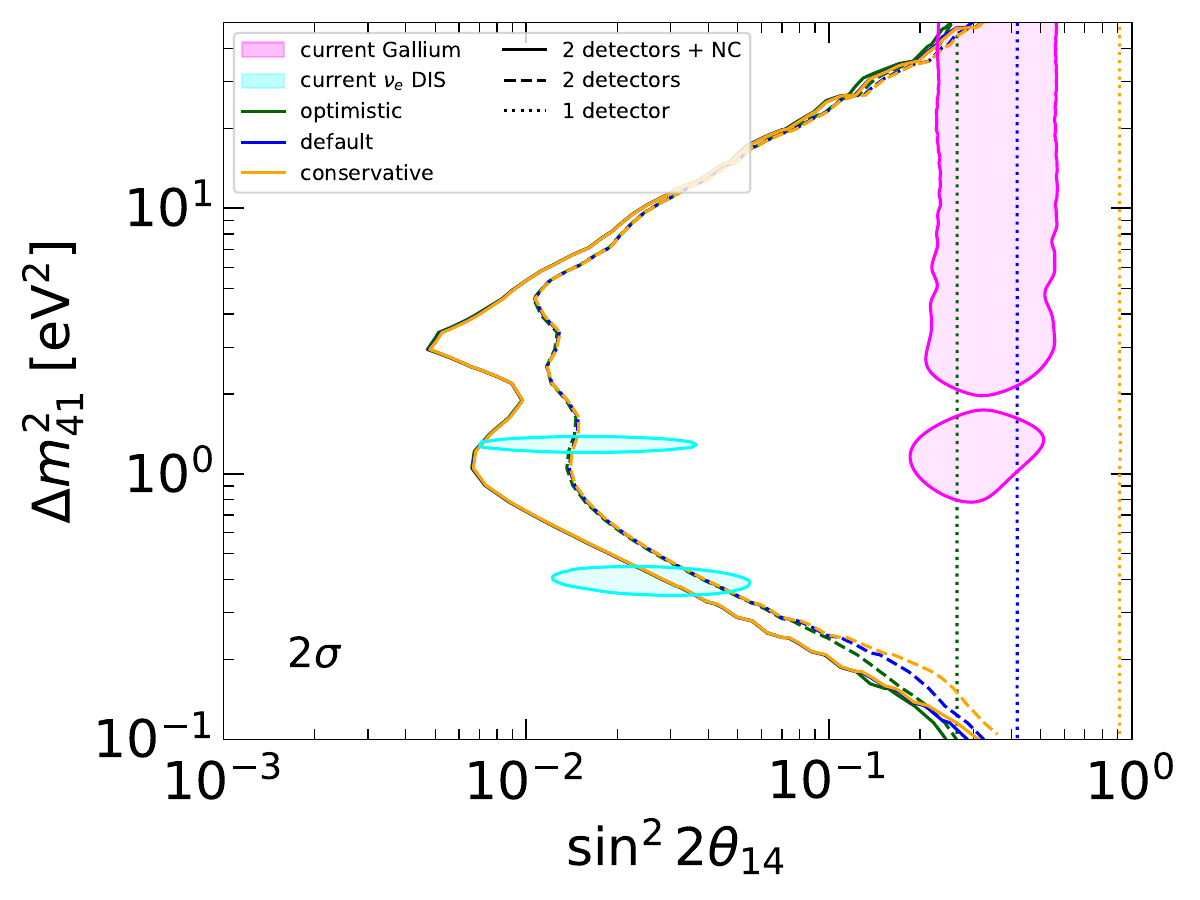}
}
\subfigure[]{ \label{fig:sq24_dm41}
\includegraphics[width=0.48\textwidth]{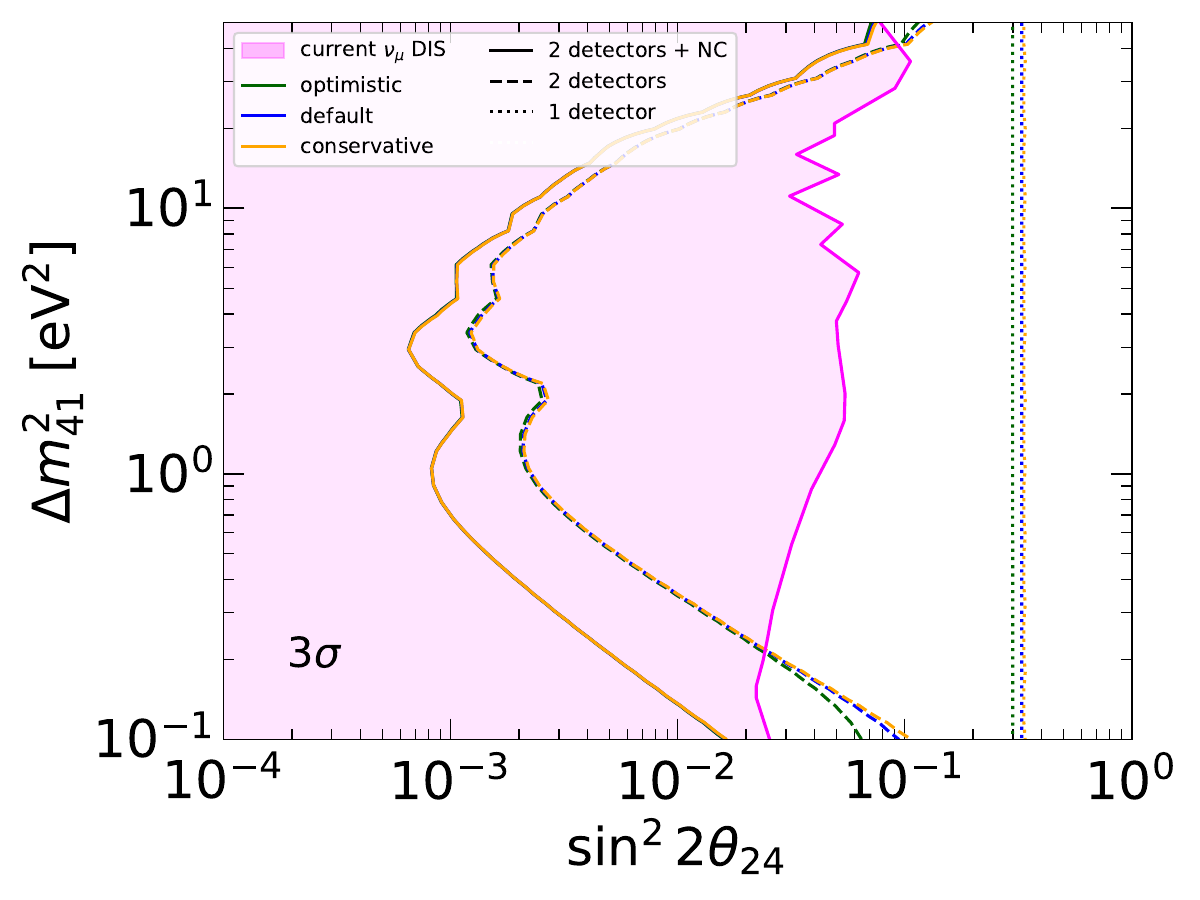}
}
    \\
\subfigure[]{ \label{fig:sqme_dm41}
\includegraphics[width=0.48\textwidth]{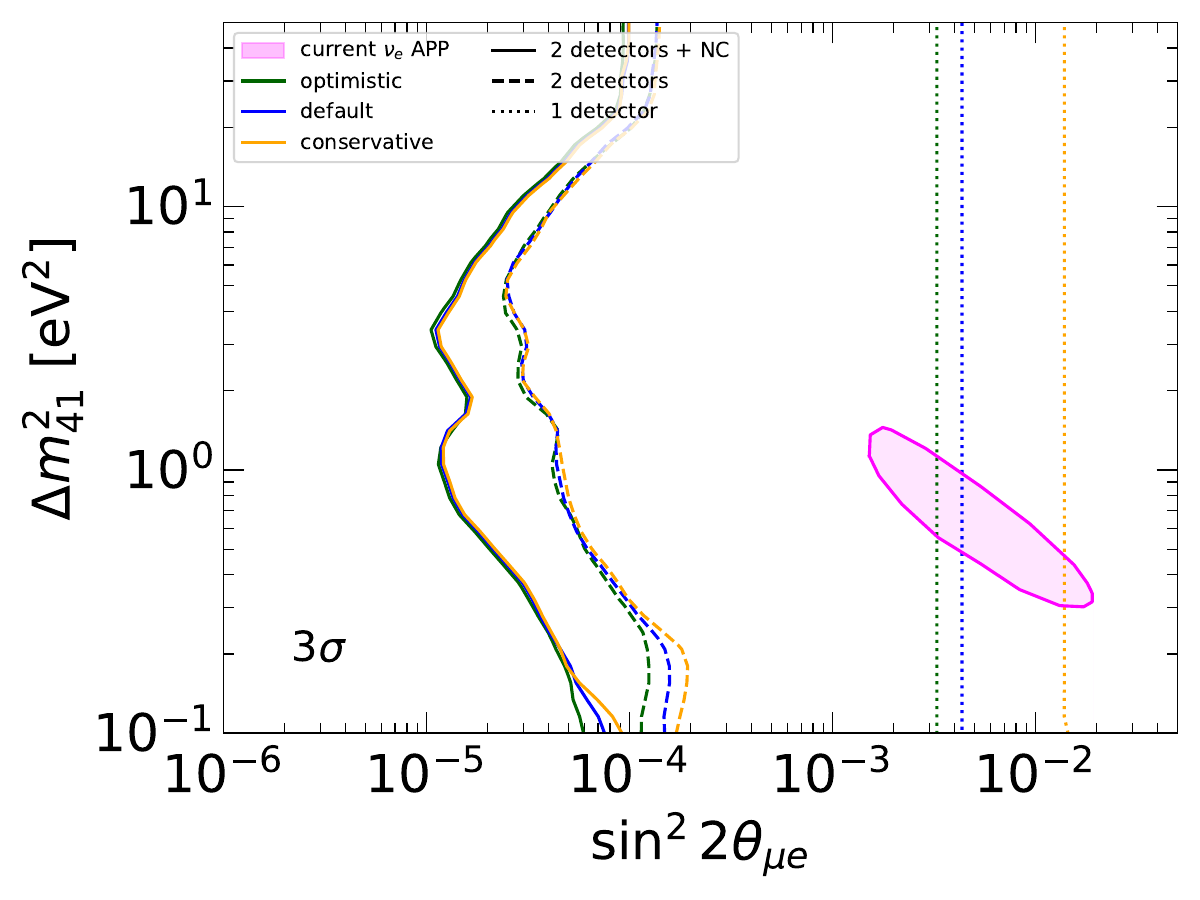}
}
\caption{Sensitivity reach of ESS$\nu$SB
in the
\subref{fig:sq14_dm41}
$\sin^2 2\vartheta_{14}$-$\Delta{m}^2_{41}$ (at $2\sigma$),
\subref{fig:sq24_dm41}
$\sin^2 2\vartheta_{24}$-$\Delta{m}^2_{41}$ (at $3\sigma$), and
\subref{fig:sqme_dm41}
$\sin^2 2\vartheta_{\mu e}$-$\Delta{m}^2_{41}$ (at $3\sigma$)
planes.
For comparison, in \subref{fig:sq14_dm41} we show also the regions preferred by the Gallium data and the regions preferred by the global $\nu_e$ disappearance analysis without Gallium data~\cite{Giunti:2022btk},
in \subref{fig:sq24_dm41} we show
the bound obtained from a global $\nu_\mu$ disappearance analysis
(the bound in Ref.~\cite{Giunti:2019aiy} updated with the latest IceCube data~\cite{IceCube:2020phf}),
and in \subref{fig:sqme_dm41} we show
the preferred region obtained from a global $\nu_e$ appearance analysis~\cite{Giunti:2019aiy}.} 
  \label{fig:sens_sterile}
\end{figure}

Figure~\ref{fig:sens_sterile} shows the results of our analysis.
As in the analyses presented in the previous Sections,
we marginalized over the standard three-neutrino oscillation parameters.
The three panels in Fig.~\ref{fig:sens_sterile}
show the ESS$\nu$SB sensitivity in the
\subref{fig:sq14_dm41}
$\sin^2 2\vartheta_{14}$-$\Delta{m}^2_{41}$,
\subref{fig:sq24_dm41}
$\sin^2 2\vartheta_{24}$-$\Delta{m}^2_{41}$, and
\subref{fig:sqme_dm41}
$\sin^2 2\vartheta_{\mu e}$-$\Delta{m}^2_{41}$
planes.
Each of them has been obtained by marginalizing
over the other active-sterile mixing angles. Again we plot the results using a single detector analysis (dotted lines), and a 2-detector analysis without (dashed lines) and with (solid lines) the inclusion of NC channels for the conservative (orange), default (blue) and optimistic (green) choices of systematic uncertainties. It is noteworthy that the inclusion of the near detector and the NC channels improves the sensitivity in all panels, while the different choices of systematic uncertainties only show some effects for very small and very large $\Delta m_{41}^2$. The reason is that for most of the range of $\Delta m_{41}^2$ plotted in the figure oscillations appear at the near detector inducing spectral distortions, while the far detector would observe only an averaged oscillation probability. Due to the different effects of the neutrino oscillation probability at the near and far detectors, the systematic uncertainties (many of which are correlated among detectors) could not cancel an oscillation effect and hence the sensitivity does not depend on the choice of uncertainty. It has been shown~\cite{Ghosh:2019zvl}, however, that the inclusion of shape uncertainties can worsen the sensitivity in this region. This type of uncertainty has not been considered here.

Figure~\ref{fig:sq14_dm41}
shows a comparison of the ESS$\nu$SB sensitivity in the
$\sin^2 2\vartheta_{14}$-$\Delta{m}^2_{41}$
plane with the allowed regions at 2$\sigma$
obtained in Ref.~\cite{Giunti:2022btk} from the analysis of the data of the
GALLEX~\cite{GALLEX:1994rym,GALLEX:1997lja,Kaether:2010ag}, SAGE~\cite{Abdurashitov:1996dp,SAGE:1998fvr,Abdurashitov:2005tb,SAGE:2009eeu}, and BEST~\cite{Barinov:2021asz,Barinov:2022wfh}
Gallium experiments which have been obtained using the traditional
Bahcall cross section model~\cite{Bahcall:1997eg}.
These Gallium allowed regions are representative of the general allowed regions which can be obtained from the Gallium data,
because other cross section models lead to similar regions~\cite{Giunti:2022btk,Berryman:2021yan,Giunti:2022xat}.
They lie at large values of $\sin^2 2\vartheta_{14}$,
which are incompatible with the requirement of
small active-sterile neutrino mixing discussed above.
One can see that ESS$\nu$SB is sensitive to the Gallium allowed region
and can rule out the 3+1 neutrino oscillation explanation of the Gallium Anomaly. In the case of optimistic uncertainties even the 1-detector analysis is capable of excluding most of the Gallium $2\sigma$ preferred region. Therefore, ESS$\nu$SB can also test the regions obtained by the Neutrino-4 collaboration~\cite{Serebrov:2020kmd}, which require similarly large mixing angles as the Gallium data. The Neutrino-4 results are, however, controversial~\cite{Danilov:2018dme,PROSPECT:2020raz,Danilov:2020rax,Giunti:2021iti}.

In Fig.~\ref{fig:sq14_dm41}
we show also the allowed regions at $2\sigma$
obtained in Ref.~\cite{Giunti:2022btk}
from the combined analysis of the data
of short-baseline $\nu_{e}$ and $\bar\nu_{e}$ disappearance experiments,
excluding the Gallium data\footnote{The shape and significance
of the allowed regions
depend on the analysis of the reactor spectral ratio data and the reactor rate data. Since at 2$\sigma$ the differences are small, we show here only one representative example,
which corresponds to RSRF(N/DB)+KI in Fig.~10 of Ref.~\cite{Giunti:2022btk}.
The interested reader is referred to the discussions in
Refs.~\cite{Giunti:2021kab,Giunti:2022btk}.}.
Since these regions lie at small values of $\sin^2 2\vartheta_{14}$,
they satisfy the requirement of
small active-sterile neutrino mixing discussed above.
As one can see from Fig.~\ref{fig:sq14_dm41},
the sensitivity of ESS$\nu$SB is not enough to probe
these $\nu_{e}$-disappearance allowed regions when considering the 1-detector analysis or the 2-detector analysis without NC channels, which does not allow probing large parts of the preferred $2\sigma$ regions. However, if the NC channels are included, the island at $\Delta{m}^2_{41} \approx 1.3 \, \text{eV}^2$ lies fully within the ESS$\nu$SB sensitivity reach, and also a large part of the island at $\Delta{m}^2_{41} \approx 0.4 \, \text{eV}^2$ can be probed.

Figure~\ref{fig:sq24_dm41}
shows a comparison of the ESS$\nu$SB sensitivity in the
$\sin^2 2\vartheta_{24}$-$\Delta{m}^2_{41}$ plane
with the current global $3\sigma$ bound, which corresponds to the bound of Ref.~\cite{Giunti:2019aiy} updated with the latest IceCube data~\cite{IceCube:2020phf}.
There is no allowed region from current data,
because the data of all
$\nu_{\mu}$ and $\bar\nu_{\mu}$
disappearance experiments are compatible with three-neutrino mixing,
without any anomaly.
From Fig.~\ref{fig:sq24_dm41},
one can see that ESS$\nu$SB can improve the current bounds for
$\Delta m_{41}^2 \gtrsim 0.3 \, \text{eV}^2$ and $\Delta m_{41}^2 \lesssim 40 \, \text{eV}^2$ using two detectors. The sensitivity can be further improved when including the NC channels.

Figure~\ref{fig:sqme_dm41}
shows the sensitivity of ESS$\nu$SB
in the
$\sin^2 2\vartheta_{\mu e}$-$\Delta{m}^2_{41}$ plane.
Let us remind that
$\sin^22\theta_{\mu e}$,
given in Eq.~\ref{eq:sme},
is the effective mixing parameter relevant for appearance experiments.
Figure~\ref{fig:sqme_dm41}
shows also the region preferred at $3\sigma$ from a global analysis of appearance experiments~\cite{Giunti:2019aiy}.
One can see that this allowed region can be tested very well in ESS$\nu$SB.

\begin{figure}
  \centering
    \includegraphics[width=0.49\textwidth]{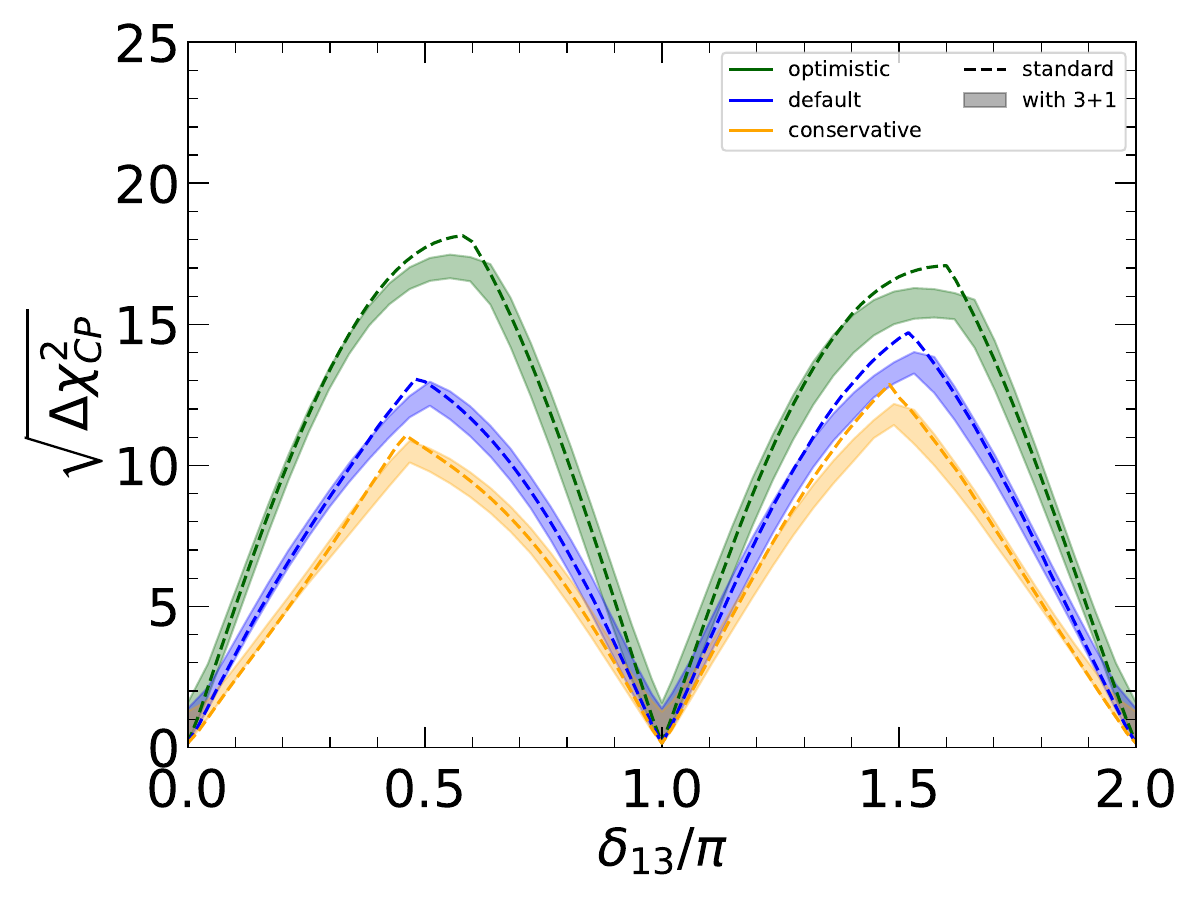}
    \caption{The impact new mixing parameters due to a light sterile neutrino could have on the determination of CP violation at ESS$\nu$SB for different choices of systematic uncertainties. The width of the bands is due to the variation of $\delta_{14}$ in the fake data. The dashed lines are the standard sensitivities obtained in Sec.~\ref{sec:SM}.
} 
  \label{fig:CP-sterile}
\end{figure}

Finally, as in the previous sections, we estimate how much the presence of a light sterile neutrino might affect the sensitivity of ESS$\nu$SB to the discovery of CP violation. We created fake data sets using as input the best fit value of the analysis in Ref.~\cite{Giunti:2022btk},
i.e. $\Delta m_{41}^2 = 1.3 \, \text{eV}^2$,
$\sin^22\theta_{14} = 0.022$. In addition we chose $\sin^22\theta_{24} = 0.022$, which is allowed from current data, but lies within the sensitivity range of ESS$\nu$SB. Note that in order to obtain some measurable effect from $\delta_{14}$, which is the CP phase of interest here, both $\theta_{14}$ and $\theta_{24}$ must be different from zero.
As in the previous sections, we generated fake data sets varying $\delta_{13}$ and also the new phase $\delta_{14}$, and then we marginalized over the CP conserving combinations of CP phases. The results of this analysis are shown in Fig.~\ref{fig:CP-sterile}. As one can see, in the case of a sterile neutrino the sensitivity to measure $\delta_{13}$ is slightly reduced. However, even at $\delta_{13}=0$ and $\delta_{13}=\pi$, there is still potential to observe CP violation at the $\sim2\sigma$ level. Note that, unlike in the case of nonunitary neutrino mixing, the bands are partially below the sensitivity curves of the standard analyses (dashed lines), instead of surrounding them. The same behavior has been observed in the cases of other long baseline experiments in Ref.~\cite{Dutta:2016glq}. It was shown that if the mixing angles are chosen to be large, the band due to the variation of $\delta_{14}$ is wider, too. Since we chose the mixing angles to be quite small, the full band lies partially below the lines for the standard sensitivity.

Note that the results presented in this section update the results of former analyses in Refs.~\cite{Blennow:2014fqa,KumarAgarwalla:2019blx,Ghosh:2019zvl}. In these references the authors used older configurations of the experiment or positions of the near detector resulting in slightly different sensitivities.

\section{Summary and conclusions}
\label{sec:conc}

We have discussed the sensitivity to CP violation and to several new physics scenarios for ESS$\nu$SB. In particular, we have shown the improvement on the sensitivity which can be obtained from a 2-detector fit and from the addition of neutral current channels. 

We have shown that ESS$\nu$SB will be able to test some of the parameter space of the NSI parameters which is preferred by the current data, as shown in the analyses of Refs.~\cite{Denton:2020uda,Chatterjee:2020kkm}. It should be noted, however, that other future experiments are expected to have similar or even better sensitivities than ESS$\nu$SB, due to the usage of larger baselines and hence larger matter effects~\cite{Ohlsson:2012kf,Miranda:2015dra,Farzan:2017xzy,Proceedings:2019qno}. In Ref.~\cite{Denton:2022pxt} the authors suggest to use DUNE to test the results from Refs.~\cite{Denton:2020uda,Chatterjee:2020kkm}. Nevertheless, the strongest bounds on several NSI parameters can be expected from future atmospheric neutrino experiments~\cite{KM3NeT:2021nnf}.

In the case of nonunitary neutrino mixing we find that ESS$\nu$SB will be able to improve some of the current bounds (e.g. $|\alpha_{21}|$), while providing comparable bounds for other parameters ($\alpha_{11}$ and $\alpha_{22}$). A great improvement can be expected with respect to the 1-detector analysis of Ref.~\cite{Chatterjee:2021xyu}. The sensitivity is similar to the one that can be expected from DUNE~\cite{Coloma:2021uhq} or T2HKK~\cite{Soumya:2021dmy,Agarwalla:2021owd}, while improving over the sensitivities of several other probes~\cite{Soumya:2021dmy,Miranda:2020syh,Gariazzo:2022evs}. 

Finally we have shown that ESS$\nu$SB has excellent sensitivity to test all of the short-baseline anomalies in the context of 3+1 neutrino mixing. Complementary sensitivities are expected at DUNE~\cite{Coloma:2021uhq}, JUNO/TAO~\cite{JUNO:2020ijm,Berryman:2021xsi,Basto-Gonzalez:2021aus}, KM3NeT~\cite{KM3NeT:2021uez}, and the SBN program at Fermilab~\cite{Machado:2019oxb}. 

Overall, we have shown that ESS$\nu$SB will be an excellent tool for the measurement of CP violation and for the exploration of several scenarios of physics beyond the standard model.


\begin{acknowledgments}
We would like to thank Salva Rosauro-Alcaraz for providing the GLoBES files for ESS$\nu$SB. C.A.T. is thankful for the hospitality at Università degli Studi dell'Aquila and LNGS where part of this work was performed.
C.G. and C.A.T. are supported by the research grant ``The Dark Universe: A Synergic Multimessenger Approach'' number 2017X7X85K under the program ``PRIN 2017'' funded by the Italian Ministero dell'Istruzione, Universit\`a e della Ricerca (MIUR). C.A.T. also acknowledges support from {\sl Departments of Excellence} grant awarded by MIUR and the research grant {\sl TAsP (Theoretical Astroparticle Physics)} funded by Istituto Nazionale di Fisica Nucleare (INFN).
\end{acknowledgments}


\bibliographystyle{apsrev4-1}
\bibliography{main}

\end{document}